\documentclass[journal]{IEEEtran}

\usepackage[dvipsnames]{xcolor}
\usepackage{fancyvrb}
\usepackage{todonotes}

\usepackage{tikz}
\usepackage{xspace}
\usepackage{mathtools}
\usepackage{subcaption}
\usepackage{amssymb}
\usepackage{booktabs}
\usepackage{pgfplotstable}
\usepackage{url}
\usepackage[inline]{enumitem}
\usepackage{listings}
\usepackage{adjustbox} 

\usepackage[binary-units]{siunitx}
\usepackage{afterpage}

\usetikzlibrary{
	fit,
	matrix,
	pgfplots.external,
	pgfplots.groupplots,
}

\usepackage{parallel}

\usepackage[author={Rev2}]{pdfcomment}

\newcommand{\etee}[4]{$#1 \underset{#4}{\xrightarrow{\mathrm{#3}}} #2$\xspace}

\newcommand{\eg}{e.g.\@\xspace}
\newcommand{\ie}{i.e.\@\xspace}
\newcommand{\PI}{PI\xspace}
\newcommand{\PIs}{PIs\xspace}
\newcommand{\PIset}{PI set\xspace}
\newcommand{\PIsets}{PI sets\xspace}
\newcommand{\etal}{~et~al.\@\xspace}
\newcommand{\cpp}{CPPs\xspace}
\newcommand{\cp}{CPP\xspace}

\newcommand{\policyImplementation}{policy implementation\xspace}
\newcommand{\policyImplementations}{policy implementations\xspace}

\newcommand{\insecureCommunication}{insecure communication\xspace}
\newcommand{\insecureCommunications}{insecure communications\xspace}

\newcommand{\InsecureCommunications}{Insecure communications\xspace}
\newcommand{\unfeasibleCommunication}{unfeasible communication\xspace}
\newcommand{\unfeasibleCommunications}{unfeasible communications\xspace}

\newcommand{\UnfeasibleCommunications}{Unfeasible communications\xspace}
\newcommand{\potentialError}{potential error\xspace}
\newcommand{\potentialErrors}{potential errors\xspace}

\newcommand{\PotentialErrors}{Potential errors\xspace}
\newcommand{\suboptimalImplementation}{suboptimal implementation\xspace}
\newcommand{\suboptimalImplementations}{suboptimal implementations\xspace}

\newcommand{\SuboptimalImplementations}{Suboptimal implementations\xspace}
\newcommand{\suboptimalWalk}{suboptimal walk\xspace}
\newcommand{\suboptimalWalks}{suboptimal walks\xspace}

\newcommand{\SuboptimalWalks}{Suboptimal walks\xspace}
\newcommand{\inadequacy}{inadequacy\xspace}
\newcommand{\Inadequacy}{Inadequacy\xspace}
\newcommand{\monitorability}{monitorability\xspace}
\newcommand{\Monitorability}{Monitorability\xspace}
\newcommand{\skewedChannel}{skewed channel\xspace}

\newcommand{\SkewedChannel}{Skewed channel\xspace}

\newcommand{\asymmetricChannel}{asymmetric channel\xspace}

\newcommand{\AsymmetricChannel}{Asymmetric channel\xspace}

\newcommand{\nonEnforceability}{non-enforceability\xspace}
\newcommand{\NonEnforceability}{Non-enforceability\xspace}
\newcommand{\outOfPlace}{out of place\xspace}
\newcommand{\OutOfPlace}{Out of place\xspace}
\newcommand{\filteredChannel}{filtered channel\xspace}

\newcommand{\FilteredChannel}{Filtered channel\xspace}

\newcommand{\filtered}{filtered\xspace}

\newcommand{\LTwo}{L2\xspace}
\newcommand{\shadowing}{shadowing\xspace}
\newcommand{\Shadowing}{Shadowing\xspace}
\newcommand{\exception}{exception\xspace}
\newcommand{\Exception}{Exception\xspace}
\newcommand{\correlation}{correlation\xspace}
\newcommand{\Correlation}{Correlation\xspace}
\newcommand{\affinity}{affinity\xspace}
\newcommand{\Affinity}{Affinity\xspace}
\newcommand{\contradiction}{contradiction\xspace}
\newcommand{\Contradiction}{Contradiction\xspace}
\newcommand{\redundancy}{redundancy\xspace}
\newcommand{\Redundancy}{Redundancy\xspace}
\newcommand{\inclusion}{inclusion\xspace}
\newcommand{\Inclusion}{Inclusion\xspace}
\newcommand{\superfluous}{superfluous\xspace}
\newcommand{\Superfluous}{Superfluous\xspace}

\newcommand{\internalLoop}{internal loop\xspace}
\newcommand{\InternalLoop}{Internal loop\xspace}
\newcommand{\alternativePath}{alternative path\xspace}

\newcommand{\AlternativePath}{Alternative path\xspace}

\newcommand{\cyclicPath}{cyclic path\xspace}

\newcommand{\CyclicPath}{Cyclic path\xspace}

\tikzstyle{root}=[draw=black,very thick,densely dotted,minimum size=16,inner sep=0,fill=white!65,circle,text=black]
\tikzstyle{int}=[draw=black,minimum size=16,inner sep=0,fill=white!65,circle,text=black]
\tikzstyle{ione}=[thick, black!50, -latex]
\tikzstyle{itwo}=[thick, densely dotted, black!50, -latex]
\tikzstyle{coef}=[text=black,anchor=north,font=\footnotesize]
\tikzstyle{tech}=[text=black,anchor=south,font=\footnotesize]
\tikzstyle{int_r}=[draw=none,minimum width = 6ex,
minimum height = 2ex, inner sep=2,fill=black!10,rectangle,text=black,rounded corners,draw=black!50,semithick]

\newcommand{\SEdge}[2]{\draw (#2.west) -- (#1.east)}
\newcommand{\REdge}[2]{\draw (#2.west) [rounded corners]-- ++(-5mm, 0mm) |- (#1.east)}
\newcommand{\NodeDistance}{40mm}

\newcommand{\NodeDistanceC}{30mm}
\newcommand{\NodeDistanceA}{25mm}
\newcommand{\NodeDistanceB}{20mm}

\newcommand{\bigcircle}[6]
{
	\draw[thick, gray, densely dotted] ([shift=(#3:#2)]#1) arc (#3:#4:#2);
	\node[circle, draw, thin, gray, minimum width=5mm, double, fill = white, xshift = {#2 * cos(#6)}, yshift = {#2 * sin(#6)}] at (#1) {#5};
}

\newcommand{\router}[2]{\node[#2] (#1) {\shortstack{\includegraphics[width=4mm]{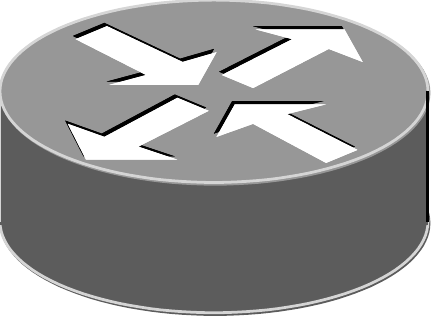}\\$#1$}}}
\newcommand{\cloud}[2]{\node[#2] (#1) {\shortstack{\includegraphics[width=8mm]{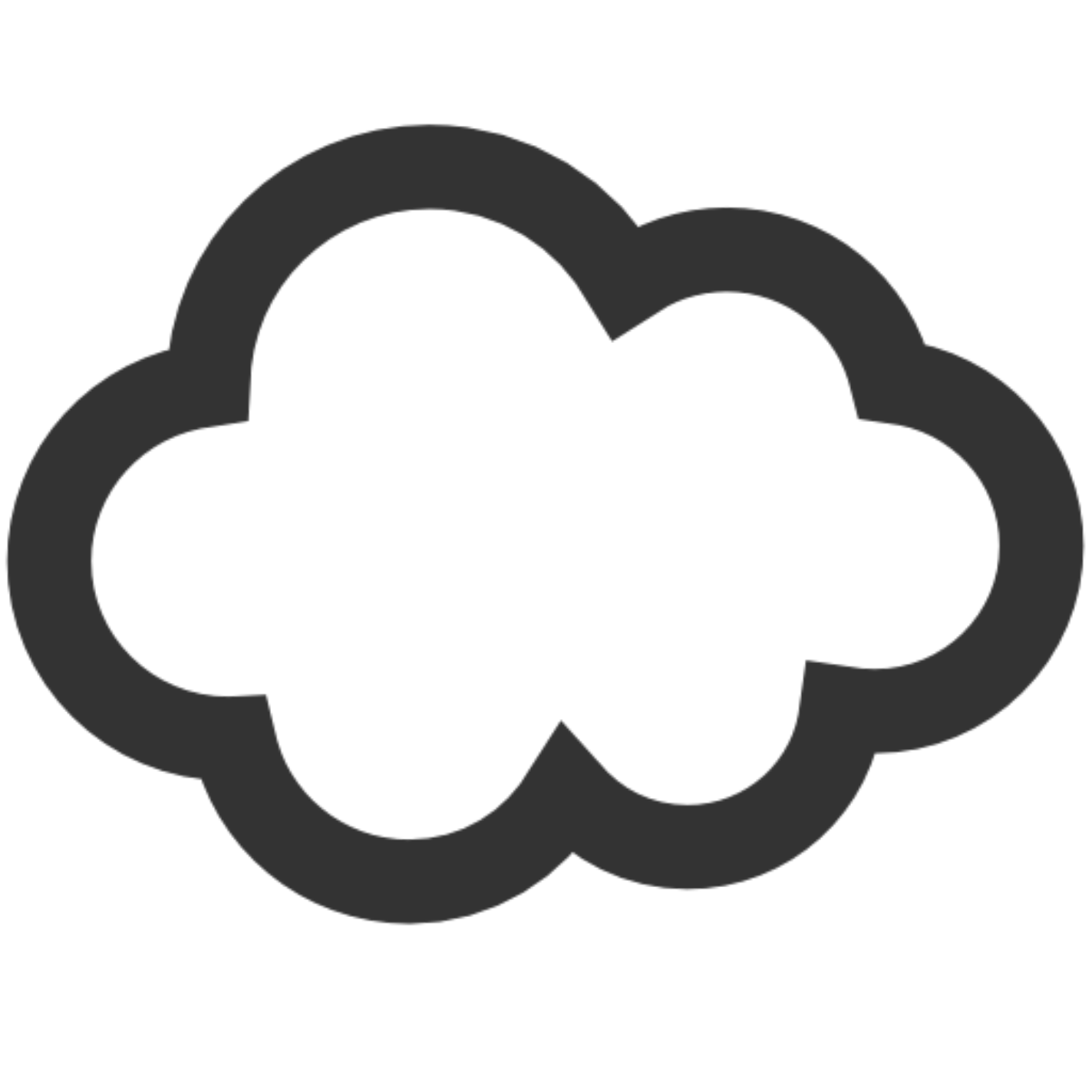}\\#1}}}
\newcommand{\client}[2]{\node[#2] (#1) {\shortstack{\includegraphics[width=4mm]{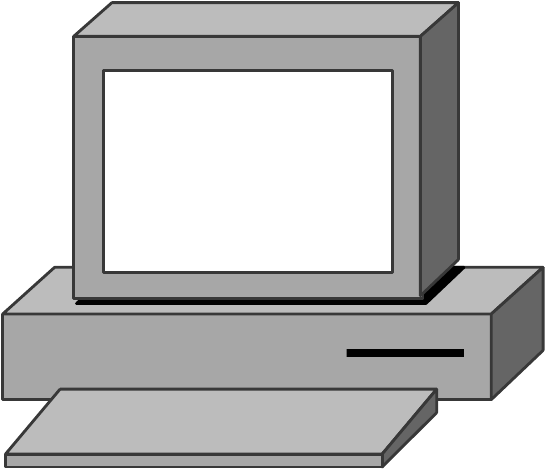}\\$#1$}}}
\newcommand{\server}[2]{\node[#2] (#1) {\shortstack{\includegraphics[width=3mm]{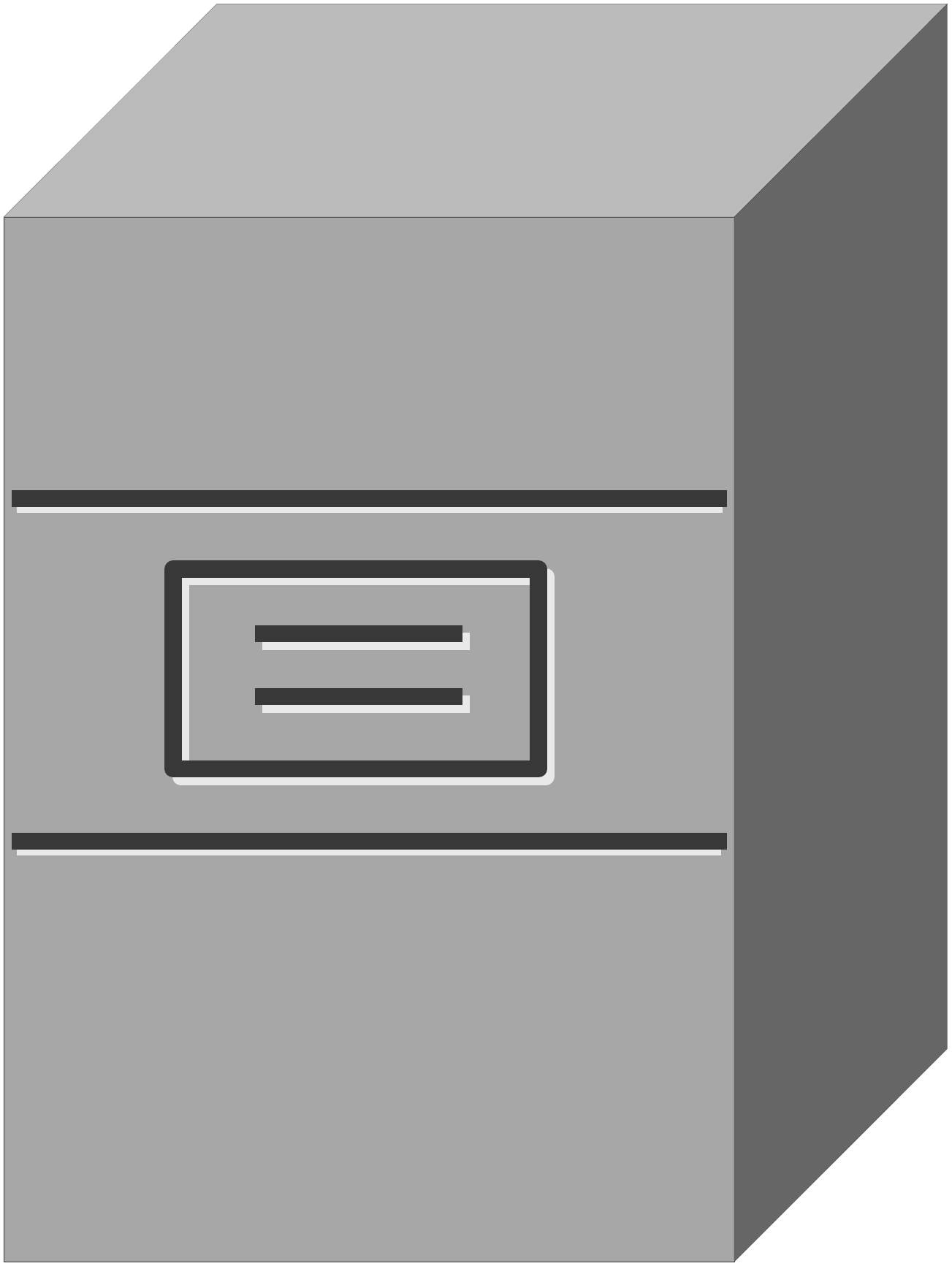}\\$#1$}}}
\newcommand{\p}[2]{\path[semithick, gray] (#1) edge (#2)}
\newcommand{\service}[2]{\node[#2,rounded corners,fill=gray!25] (#1) {$#1$}}

\newcommand{\clientA}[2]{\node[#2] (#1) {\shortstack{\includegraphics[width=4mm]{clientg.pdf}\\$#1$}}}
\newcommand{\serverA}[2]{\node[#2] (#1) {\shortstack{\includegraphics[width=3mm]{serverg.pdf}\\$#1$}}}
\newcommand{\routerA}[2]{\node[#2] (#1) {\shortstack{\includegraphics[width=4mm]{routerg.pdf}\\$#1$}}}
\newcommand{\pA}[3]{\draw[semithick, gray, #3] (#1) -- (#2)}

\title{Classification and Analysis of Communication Protection Policy Anomalies}
\author
{		Fulvio Valenza,
		Cataldo Basile,
		Daniele Canavese and
		Antonio Lioy
\thanks{This work has been partly supported by the SECURED and SHIELD project (grant agreements no.
	611458 and 700199), co-funded by the European Commission. 
The authors gratefully thank Dr. Marco Torchiano and Dr. Gabriella Dardanoni for their feedback on the empirical study.
}
\thanks{F. Valenza, C. Basile, D. Canavese, and A. Lioy are with the Politecnico di Torino, Dip. Automatica e Informatica, Torino (Italy).
		Email: \{first.last\}@polito.it, F.~Valenza is also with the CNR-IEIIT, Torino (Italy); e-mail: fulvio.valenza@ieiit.cnr.it.}
}

\begin{document}

\maketitle

\begin{abstract}
This paper presents a classification of the anomalies that can appear when designing or implementing communication protection policies.
%
%
Together with the already known intra- and inter-policy anomaly types, we introduce a novel category, the inter-technology anomalies, related to security controls implementing different technologies, both within the same network node and among different network nodes.
Through an empirical assessment, we prove the practical significance of detecting this new anomaly class.
Furthermore, this paper introduces a formal model, based on first-order logic rules that analyses the network topology and the security controls at each node to identify the detected anomalies and suggest the strategies to resolve them.
This  formal model has manageable computational complexity and its implementation has shown excellent performance and good scalability.

\end{abstract}

\begin{IEEEkeywords}
protection policy, policy anomalies, policy conflicts, network security.
\end{IEEEkeywords}

\section{Introduction}
\label{sec:Introduction}
Enforcing  network and communication security in a computer system is a very complex and sensitive task. 
Security administrators have a hard job to accomplish since it requires very specific skills and a high level of competence. 
On one hand, several studies confirm the administrators' responsibilities in many security breaches and breakdowns. 
Wool showed that most of the analysed firewalls contain several problematic policies, such as very lax rules \cite{Wool2010}.
The 2015 Data Breach Investigations Report states that about 60\% of the security breaches are due to internal staff errors \cite{Verizon2015}.
On other hand, administrators often have basic or no tool support to debug the security controls' configurations and check if the enforced policy is compliant with the high-level security requirements.

Configuration analysis methods, often addressed in literature, have seldom been incorporated into industrial tools for various reasons.
For instance, some analysis methods may be challenging to use because their computational complexity make them inapplicable or some others make assumptions that are far from reality. The only notable exception is the detection of packet filter anomalies classified by Al-Shaer \cite{Al-Shaer2005}, which is available in some Cisco routers \cite{cisco}.

The main contribution of this paper is a formal model to assist security administrators in configuring and validating security policies. 
We focus on \emph{communication protection policies} (\cpp), which determine how to protect assets (such as private user data and corporate intellectual properties) when they are transferred over computer networks. 
\cpp are very sensitive, as incorrect implementations can result in information disclosure that can cause violations of the users' privacy, intellectual properties and, often, huge monetary losses.
Moreover, CPP are particularly difficult to manage and enforce since the probability of introducing them increases with the number of entities involved in the implementation, as well as with the size and complexity of the network to configure \cite{AlIPsec,Wool2010}. Therefore, there is a non-negligible risk of mistakes, such as introducing contradicting configurations and redundant channels.

\cpp originate from legal (e.g. the EU privacy law \cite{eulaw}) and business security requirements. 
They are often expressed with high-level directives specifying the security properties that the communication must guarantee (\eg confidentiality and integrity).
In some cases, such as companies that host services or provide cloud-based resources, \cpp may be very elaborated and complex.
IT managers generally interact with business units to determine what to protect.
However, \cpp are enforced by a plethora of security controls implementing different protocols (\eg IPsec, TLS, WS-Security) that operate at different layers of the OSI stack.
Service administrators have to enable channel protection protocols on the services they manage. 
Finally, \emph{network and security administrators} have to decide the resources accessible through secure protocols and the data that need protection.
Consequently, they have to map the selected resources and data to network entities and traffic to protect, decide the OSI layer where to apply the protection, select whether to create end-to-end channels or tunnels, when to enable wireless protection, and so on.

The approach we follow in this paper is to detect and show to the administrators the \emph{anomalies}, which are the presence of redundant or conflicting configuration rules. 
Anomalies often reveal human errors and thus deserve the explicit attention of the administrators.
%
Detecting anomalies helps in daily \cpp' management activities.
On one hand, when designing a \cp implementation, debugging the policy before implementing it can speed-up the deployment time and avoid critical problems (\emph{\cp design validation}). With our approach, the administrator has to provide in input the communications he wants to protect, the security properties to be enforced and the technology he wants to use. The model we developed will output the detected anomalies, such as redundant, insecure, non-enforceable and filtered communications, and propose resolutions that significantly help in reducing human errors.
On the other hand, our approach can help in assuring that the current \cp implementation respects its intended semantic.
In the \emph{\cp implementation analysis}, the administrator provides the network topology and the configuration of all the protection controls  to get as output the anomalies, explanations, and suggested remediation actions.

This paper is structured as follows.
Section~\ref{sec:Contributions} lists our contributions to the current state-of-the art.
Sections~\ref{sec:Background} and~\ref{sec:Example} informally introduce our approach by presenting some background and a motivating example.
Sections~\ref{sec:Structure} and~\ref{sec:Analysis} are the core of this paper and describe the formal structures and the formulas of our model. Section~\ref{sec:Graph} presents the graphical notations for reporting the anomalies.
Section~\ref{sec:Validation} contains the complexity and performance analysis of the presented approach, together with an empirical study.
Finally, Sections~\ref{sec:RelatedWorks}~and~\ref{sec:Conclusions} contain the related works and the conclusions.


\section{Contributions}
\label{sec:Contributions}

Our work pushes forward the state-of-the-art in several directions.
The main contribution is the identification of nineteen types of anomalies that may happen when implementing a \cp.
Six anomaly types are already known \cite{AlIPsec}, but all the others are our original contribution\footnote{We presented an embryonic and incomplete set of anomalies in a previous work \cite{crisis}. {Note that the ``superfluous'' keyword was already used in another work about firewall conflicts to denote redundant and shadowed rules. It is a completely different meaning, as it will be evident from this treatment \cite{Amina}.}}.
%

The anomalies we identify arise in the configuration of a single security control (\emph{intra-policy}), between controls of the same type displaced at different network nodes (\emph{inter-policy}), and, our novel contribution, among security controls implementing different technologies, within a single network node or among different network nodes (\emph{inter-technology}).
We focus on communication protection controls that work at four network layers: data link, network, session\footnote{Protections at transport layer, such as TLS, are sometimes associated to the session layer as they work on top of the TCP/UDP protocols. We do not want to enter a philosophical diatribe as, for our purposes, the important thing is the order of encapsulation of the different protections.}, and application.
As an example, inter-policy anomalies may appear between the configurations of two IPsec gateways, while inter-technology anomalies may arise between the IPsec and TLS configurations implemented at the same network node.
Moreover, we also identify communications that are intrinsically insecure or non-enforceable by the target security controls.

Anomalies are detected by means of a formal model that takes as input the network description, nodes information and {the security controls' configurations and the communications to secure, during a \cp design validation, or the communications actually secured, during a \cp implementation analysis}.
Information in input become part of a knowledge base explored with a set of first-order logic (FOL) formulas to identify and report the detected anomalies to the administrators.

Anomalies have been categorized according to two different classifications: an \emph{effect-based} taxonomy and a \emph{information-centric} one.
The first classification divides the anomalies into five macro-categories describing the effects that they have on the network:
{\begin{enumerate*}
		\item insecure communications;
		\item unfeasible communications;
		\item potential errors;
		\item suboptimal communications;
		\item suboptimal walks.
\end{enumerate*}}
The second one is based on the information needed to be analysed for detecting the anomalies. {It divides the anomalies into three classes:
	\begin{enumerate*}
		\item anomalies in a single communication channel;
		\item anomalies between secure channels that start at or end on the same node;
		\item anomalies that are only evident if the full network information (nodes and topology) and high-level security requirements are considered.
\end{enumerate*}}

Having introduced several new kinds of anomaly, we posed ourselves several questions regarding the impact of our work:

\begin{enumerate}
	\item is detecting these anomalies important and helpful to improve the security of the current IT infrastructures?
	\item are these anomalies actually introduced by the administrators when they implement their policies?
	\item is it computationally feasible to identify these anomalies in large networks?
\end{enumerate}

In order to answer the first question, for each anomaly we  present the possible consequences on the network and some ways to resolve it for reducing the security impacts on the short and long period (see Section~\ref{sec:Analysis}).
To answer the second question, we prepared an empirical experiment where three categories of administrators (experts, intermediate and beginners) were asked to configure a set of \cpp in a sample network. We noticed that several of the newly introduced anomalies appeared (see Section~\ref{sec:Empirical}).
And finally, to answer the last question, we implemented, and tested in several different scenarios, a tool making use of DL (description logic) ontologies and custom Java-based reasoning rules (see Section~\ref{sec:Performances}).

\section{Background}
\label{sec:Background}
%
A \emph{communication} is any directional data exchange between two network entities. A \emph{secure communication} is a communication `adequately' protected, that is it fully satisfies a set of security requirements.
In this context, the security requirements concern three \emph{security properties}: header integrity, payload integrity and (payload) confidentiality.


A \emph{channel} is a directional data exchange between two nodes protected with some security properties (a \emph{secure} channel) or none (an \emph{insecure} channel).
Logically, a secure channel is an association between a source, where the security properties are applied, and a destination, where the security properties are removed or verified.
A communication can be thought as a stack of several (secure and/or insecure) channels.
For example, an end-to-end TLS communication consists of a single secure channel.
However, more complex scenarios exist.
For instance, a communication between two hosts in separate networks connected via an IPsec site-to-site VPN is modelled with two channels: an insecure one between the end-points and an IPsec secure channel between the VPN terminators.


In the real world, the secure communications are defined by using a set of configuration settings containing several low-level details.
For instance, the configuration of a TLS server contains detailed information about the supported cipher-suites.
However, during the design and policy analysis phases, this level of granularity is usually not needed.
For our purposes, a secure channel can be represented by specifying:
\begin{enumerate*}
	\item the source and destination entities (they can be network nodes or direct references to an entity lying at a particular OSI layer such IP addresses and URIs);
	\item the security protocol to use (our model can be easily extended to new protocols and can support a wide array of technologies at different OSI layers);
	\item the required security properties;
	\item the crossed gateways and the traffic to protect (meaningful only in case of tunnels).
\end{enumerate*}
We name \emph{\policyImplementation} (or \emph{\PI} for short) this formal representation of a channel. 
Note that since a channel is directional, a \PI is directional too. 
This means that, to create a complete request-reply connection, we need at least two \PIs. More information on this subject is provided in Section~\ref{sec:Structure}.

{We call a \emph{\PIset} a group of \policyImplementations that belongs to the same node and use the same technology. For instance, a particular server supporting IPsec and SSH will have two \PIsets, one for each protocol. We will assume without loss of generality that the \policyImplementations in the same \PIset are ordered according to their priority\footnote{The work \cite{Basile12} proved that any policy represented as a set of rules can be expressed as an equivalent policy where rules are ordered by their priority.}.}
Note that our analysis uses other additional sources of information:

\begin{itemize}
	\item \emph{network reachability data}, as the configurations of filtering controls and NAT devices must be available to determine if the channels can be actually established (\eg to check if a channel is not dropped by a firewall);
	\item \emph{supported security protocols} (at various OSI levels), for guaranteeing that it is possible to establish the secure channels;
	\item \emph{supported cryptographic algorithms}, as some cipher-suites might not be available when actually deploying a \PI on the installed security controls.
\end{itemize}

Finally, we will refer to the \emph{network topology} as a graph where its nodes are potential channel end-points (both sources and destinations) and its edges are physical or virtual connections between them.

\section{Motivating example}
\label{sec:Example}

\begin{figure}[t]
	\centering
	{

\begin{tikzpicture}[node distance=12mm, inner sep=0.5mm]
\footnotesize
\cloud{Internet}{};

\router{g_{a1}}{above left of=Internet, xshift=-5mm, yshift=8mm};
\client{c_{a2}}{left of=g_{a1}};
\client{c_{a1}}{above left of=g_{a1}};
\client{c_{a3}}{below left of=g_{a1}};
\p{Internet}{g_{a1}};
\p{g_{a1}}{c_{a1}};
\p{g_{a1}}{c_{a2}};
\p{g_{a1}}{c_{a3}};

\router{g_{b1}}{below left of=Internet, xshift=-5mm, yshift=-8mm};
\client{c_{b1}}{above left of=g_{b1}};
\client{c_{b2}}{below left of=g_{b1}};
\p{Internet}{g_{b1}};
\p{g_{b1}}{c_{b1}};
\p{g_{b1}}{c_{b2}};
\p{g_{b1}}{c_{b2}};

\router{g_{c1}}{right of=Internet, xshift=4mm};
\router{g_{c3}}{right of=g_{c1},xshift=2mm, yshift=-12mm};
\router{g_{c2}}{right of=g_{c1},xshift=2mm, yshift=12mm};
\server{s_{c1}}{above right of=g_{c2}};	
\server{s_{c2}}{right of=g_{c2}};
\service{db}{above right of=s_{c1}};
\service{web_1}{right of=s_{c1}};
\client{c_{c2}}{right of=g_{c3}};
\service{web_2}{right of=s_{c2}};
\client{c_{c1}}{above right of=g_{c3}};
\client{c_{c3}}{below right of=g_{c3}};
\p{Internet}{g_{c1}};
\p{g_{c1}}{g_{c2}};
\p{g_{c1}}{g_{c3}};
\p{g_{c2}}{s_{c1}};
\p{g_{c2}}{s_{c2}};
\p{g_{c3}}{c_{c1}};
\p{g_{c3}}{c_{c2}};
\p{g_{c3}}{c_{c3}};
\p{g_{c2}}{g_{c3}};
\p{s_{c1}}{db};
\p{s_{c1}}{web_1};
\p{s_{c2}}{web_2};

\node[fit = (c_{a1}) (c_{a3})] (a) {};
\node[fit = (c_{b1}) (c_{b2})] (b) {};
\node[fit = (s_{c1}) (c_{c3})] (c) {};

\bigcircle{a.center}{16mm}{-70}{70}{A}{50}
\bigcircle{b.center}{16mm}{-70}{70}{B}{-50}]
\bigcircle{c.center}{32mm}{110}{250}{C}{120}]

\end{tikzpicture}

	}
	\caption{A simplified network scenario.}
	\label{fig:SampleNetwork}
\end{figure}
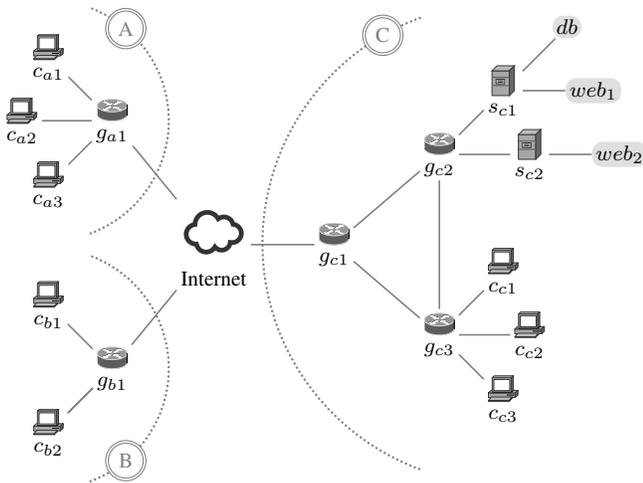

Before formally tackling the analysis of the \PI anomalies, we  begin our discussion by considering the simplified network scenario in Fig.~\ref{fig:SampleNetwork}.
The diagram shows a main corporate network ($C$) and two branch networks ($A$ and $B$).
The three networks are connected via the Internet and consist of a number of security gateways (denoted by $g$) that mediate the communications between the servers ($s_{c1}$ and $s_{c2}$) and the clients (indicated by $c$).
The server $s_{c1}$ hosts two services ($web_1$ and $db$), while $s_{c2}$ hosts only one web service ($web_2$).

We will use the informal notation \etee{s}{d}{\;\;t\;\;}{pi,c} to indicate a \PI that establishes a channel from the source $s$ to the destination $d$ using the technology $t$ to enforce some security properties. In this simplified example we will only take into account two security properties, (payload) confidentiality and payload integrity, denoted by the symbols  $c$ and $pi$, respectively. For instance, the \PI \etee{a}{b}{IPsec}{pi} indicates an IPsec connection with integrity (but not confidentiality), from $a$ to $b$.

\begin{figure}[ht]
	\centering
	\scalebox{0.8}
	{

\begin{tikzpicture}[node distance = 6mm, semithick, draw = gray]
\node (Inadequate) {\inadequacy};
\node[below = of Inadequate.west, anchor = west]  (Monitorable) {\monitorability};
\node[below = of Monitorable.west, anchor = west]   (Skewed) {\skewedChannel};
\node[below = of Skewed.west, anchor = west] (Asymmetric) {\asymmetricChannel};

\node[below = of Asymmetric.west, anchor = west] (NonEnforceable) {\nonEnforceability};
\node[below = of NonEnforceable.west, anchor = west] (Misplaced) {\outOfPlace};
\node[below = of Misplaced.west, anchor = west] (Filtered) {\filtered};
\node[below = of Filtered.west, anchor = west] (L2) {\LTwo};

\node[below = of L2.west, anchor = west] (Shadowing) {\shadowing};
\node[below = of Shadowing.west, anchor = west] (Exception) {\exception};
\node[below = of Exception.west, anchor = west] (Correlation) {\correlation};
\node[below = of Correlation.west, anchor = west] (Affine) {\affinity};
\node[below = of Affine.west, anchor = west] (Hide) {\contradiction};

\node[below = of Hide.west, anchor = west] (Redundancy) {\redundancy};
\node[below = of Redundancy.west, anchor = west] (Inclusion) {\inclusion};
\node[below = of Inclusion.west, anchor = west] (Superfluous) {\superfluous};
\node[below = of Superfluous.west, anchor = west] (IntraChannel) {\internalLoop};

\node[below = of IntraChannel.west, anchor = west] (Alternative) {\alternativePath};
\node[below = of Alternative.west, anchor = west] (Loop) {\cyclicPath};

\node[fit = (Inadequate) (Monitorable)   (Skewed) (Asymmetric)  ] (UnsecureBox) {};
\node[fit = (NonEnforceable)  (Filtered) (L2)  (Misplaced) ] (UnfeasibleBox) {};
\node[fit = (Redundancy) (Inclusion)  (Superfluous) (IntraChannel) ] (SubotimalImpBox) {};
\node[fit = (Shadowing)  (Exception)  (Correlation) (Affine)(Hide)] (DiscrepantBox) {};
\node[fit =  (Loop) (Alternative)] (WalkBox) {};

\node[left =  \NodeDistance of UnsecureBox, anchor = west] (Unsecure) {\shortstack{insecure \\communications}};
\node[left = \NodeDistance of UnfeasibleBox, anchor = west] (Unfeasible) {\shortstack{unfeasible \\communications}};
\node[left = \NodeDistance of SubotimalImpBox, anchor = west] (SubotimalImp) {\shortstack{suboptimal\\implementations}};
\node[left = \NodeDistance of DiscrepantBox, anchor = west] (Discrepant) {\potentialErrors};
\node[left = \NodeDistance of WalkBox, anchor = west] (Walk) {\suboptimalWalks};

\node[fit =(Unsecure) (Unfeasible) (Discrepant)  (SubotimalImp)   (Walk) ] (AnomaliesBox) {};

\node[left = 30mm of AnomaliesBox, anchor = west] (Anomalies) {anomalies};

\REdge{Anomalies}{Unsecure};
\REdge{Anomalies}{Unfeasible};
\REdge{Anomalies}{Discrepant};
\REdge{Anomalies}{SubotimalImp};
\REdge{Anomalies}{Walk};

\node[fit = (IntraChannel)  (Loop) (Alternative)] (WalkBox) {};

\REdge{Walk}{Loop};
\REdge{Walk}{Alternative};

\REdge{Discrepant}{Shadowing};
\REdge{Discrepant}{Exception};
\SEdge{Discrepant}{Correlation};
\REdge{Discrepant}{Affine};
\REdge{Discrepant}{Hide};

\REdge{Unsecure}{Inadequate};
\REdge{Unsecure}{Monitorable};
\REdge{Unsecure}{Skewed};
\REdge{Unsecure}{Asymmetric};

\REdge{Unfeasible}{NonEnforceable};
\REdge{Unfeasible}{Filtered};
\REdge{Unfeasible}{L2};
\REdge{Unfeasible}{Misplaced};

\REdge{SubotimalImp}{Inclusion};
\REdge{SubotimalImp}{Redundancy};
\REdge{SubotimalImp}{Superfluous};
\REdge{SubotimalImp}{IntraChannel};

\end{tikzpicture}

	}
	\caption{Effect-based taxonomy.}
	\label{fig:MacroCategoriesTaxonomy}
\end{figure}
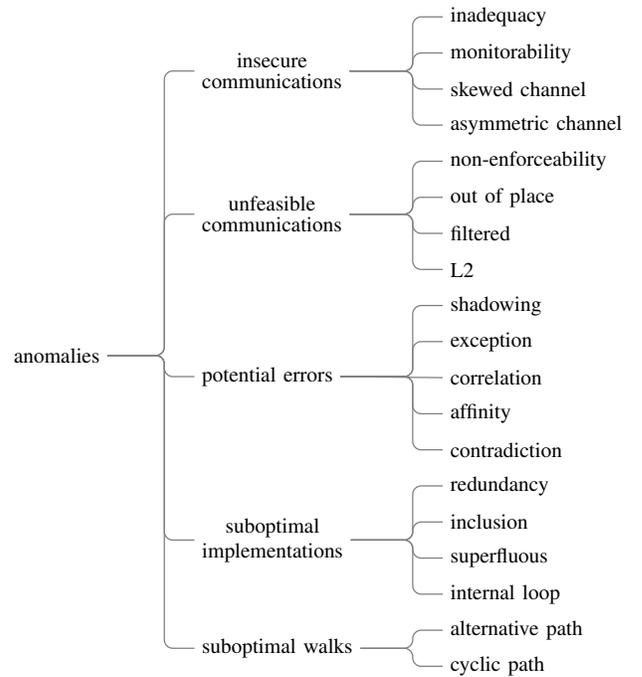

For the sake of clarity, we grouped the anomalies into five macro-categories, as shown in Fig.~\ref{fig:MacroCategoriesTaxonomy}.
These macro categories will be briefly described in the next paragraphs.

\subsection{\InsecureCommunications}

We have an \emph{\insecureCommunication} when its security level is less than the expected one.
For instance, a channel that does not satisfy the minimum security level specified in the corporate policy generates an \emph{\inadequacy} anomaly. 
We have this anomaly if the IT managers require that `all the data crossing the Internet must be encrypted' and a security administrator creates the \policyImplementation \etee{c_{a1}}{s_{c1}}{TLS}{pi}.

Another case of insecure communication arises when the security requirements are respected but we have a communication consisting of more than one channel (\eg remote-access).
In this case, the nodes at the channel junctions can `see' the exchanged data, thus lowering the security of the connection and creating a \emph{\monitorability} anomaly. For instance, the \PIs \etee{s_{c1}}{g_{c1}}{IPsec}{c, pi} and \etee{g_{c1}}{c_{a1}}{IPsec}{c, pi}  create a form of logical communication between $s_{c1}$ and $c_{a1}$ composed of a sequence of two channels interconnected through $g_{c1}$.
This means that, even if everything is encrypted, $g_{c1}$  reads the payload because it decrypts and encrypts the exchanged data (note that this may be the intended behaviour).

\begin{figure}[ht]
	\centering
	\begin{tikzpicture}
	\draw[very thick, ->] (0, 0.25) -- ++(0, -1.6) node[below] {$g_{c3}$};
	\draw[very thick, ->] (2, 0.25) -- ++(0, -1.6) node[below] {$g_{c1}$};
	\draw[very thick, ->] (4, 0.25) -- ++(0, -1.6) node[below] {$g_{a1}$};
	
	\draw[semithick, double, gray, -latex] (0, 0) -- ++(4, -0.2);
	\draw[semithick, gray, -latex] (4, -0.25) -- ++(-2, -0.3);
	\draw[semithick, densely dashed, gray, -latex] (2, -0.6) -- ++(2, -0.5);
	
	\draw[semithick, double, gray, -latex] (5, -0.25) -- ++(1, 0) node[black, right] {double tunnel};
	\draw[semithick, gray, -latex] (5, -0.75) -- ++(1, 0) node[black, right] {single tunnel};
	\draw[semithick, densely dashed, gray, -latex] (5, -1.25) -- ++(1, 0) node[black, right] {no tunnel};
	\end{tikzpicture}
	\caption{Diagram of the \skewedChannel between $g_{c3}$, $g_{c1}$, $g_{a1}$.}
	\label{fig:TrellisDiagram}
\end{figure}
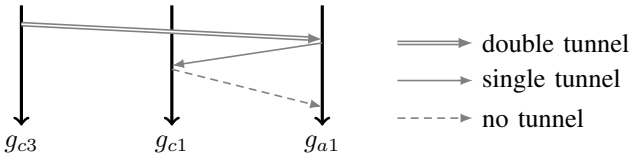

Another kind of \insecureCommunication, more subtle but potentially catastrophic, can occur with a wrong tunnel overlapping that removes the confidentiality in a part of the communication and produces a \emph{\skewedChannel} anomaly (a super-set of the Hamed\etal's overlapping anomalies \cite{AlIPsec}). For example, a security administrator can create a tunnel \etee{g_{c3}}{g_{a1}}{IPsec}{c} and another one with \etee{g_{c3}}{g_{c1}}{IPsec}{c} (note that the latter tunnel is `included' in the first one). The trellis diagram in Fig.~\ref{fig:TrellisDiagram} helps to graphically visualize the problem.

When $g_{c3}$ sends some data, it encapsulates the information in two tunnels. Hence when $g_{a1}$ receives the data, it removes the external tunnel encapsulation, but cannot remove the internal one, so $g_{a1}$ sends the data back to $g_{c1}$ which, in turn, removes the last tunnel. Finally, $g_{c1}$ sends the data to $g_{a1}$  with no protection, thus exposing the communication content to a sniffing attacker.

In the real world, most of the connections are bidirectional, since a request usually requires a reply. It may be the case that the request channel has a different security level from the reply one, generating an \emph{\asymmetricChannel} anomaly. This is not necessarily an issue, but it could be useful to report this inconsistency, so that administrators can check if the security control configurations reflect the intended network behaviour.

\subsection{\UnfeasibleCommunications}

An \emph{\unfeasibleCommunication} is a communication that cannot be established due to a hard misconfiguration.
These anomalies are very severe since they completely prevent any data exchange.
The simplest example of an \unfeasibleCommunication is when the security administrators design a \PI with a technology not supported by an end-point or a security level too high to be enforced by the available cipher-suites.
We call this situation a \emph{\nonEnforceability} anomaly.
For instance, the policy implementation \etee{web_2}{db}{TLS}{c} becomes non-enforceable if the service administrators did not install a TLS module on $s_{c2}$ (where $web_2$ resides). This \PI must obviously be  deployed on the source endpoint $s_{c2}$, however, if the node containing this \PI is not $s_{c2}$, we have generated another problem, an \emph{\outOfPlace} anomaly.

We have also a hindered connection if the packets of a channel are dropped by a firewall that lies on the path between the source and destination, thus producing a \emph{\filteredChannel} anomaly.

Firewalls and bad server configurations are not the only causes of an \unfeasibleCommunication.
There are also technological incompatibilities between wired and wireless protocols when performing security at level 2 (data link) of the OSI stack.
For example, if we choose to create a secure channel using the WPA2 technology, we must be sure that the network frames only cross wireless-enabled nodes.
If one or more crossed devices are wired-only, then we have an \emph{\LTwo anomaly}.

\subsection{\PotentialErrors}

\emph{\PotentialErrors} are a class of anomalies  where the original intent of the administrators is unclear. Hence their resolution requires a full human inspection.
When working with a large group of \PIs, an administrator can create a \PI that meddles all the traffic of another one that  has different security properties.  For instance, the \PI \etee{c_{a1}}{web_1}{TLS}{pi} hides \etee{c_{a1}}{web_1}{TLS}{c}, if the first one has a higher priority. Since the first one shadows the second one we call this anomaly a \emph{\shadowing} anomaly. If the second \PI instead has a higher priority, we have an \emph{\exception} anomaly. Exceptions are useful and are typically exploited by administrators to express an `all but one' rule, but we report them for verification.

Another kind of \potentialError is when we have two \PIs with the same technology and with the source and destination on the same node.
This situation can lead to an ambiguity, since sometimes a piece of data can match multiple \PIs, hence making the intended protection level unclear.
For example, \etee{web_2}{s_{c1}}{TLS}{c} and \etee{s_{c2}}{db}{TLS}{c, pi} are ambiguous since a packet from $web_2$ to $db$ can match both \PIs.
We call this problem a \emph{\correlation} anomaly. Analogously, we will have an \emph{\affinity} anomaly between two \PIs that use  different technologies but have the source and destination on the same nodes.

{
	Finally, we have a \emph{\contradiction} anomaly when two \PIs respectively express that the same communication should be protected and not protected.
	For instance, let suppose that an IT manager defines a policy where `all the traffic for the Internet must be inspected' and a security manager enforces the encryption of the traffic exchanged between $c_{a1}$ and $s_{c1}$ via the \PI \etee{c_{a1}}{s_{c1}}{IPsec}{c, pi}.
	This leads to a \contradiction, since the policy requires that the data for the Internet should be monitorable, but the \PI encrypts them.
}

\subsection{\SuboptimalImplementations}

\emph{\SuboptimalImplementations} arise when one or more \PIs can decrease the network throughput by producing some overhead in the nodes. Their existence is usually not problematic, but their resolution can be beneficial since it improves the network performance and makes the \PIs less vulnerable to DoS attacks.
The simplest kind of \suboptimalImplementation occurs when an administrator deploys a \PI that makes another one useless, as the first one can secure the communication at the same or  a higher level of the second, with a more effective protection (\eg stronger encryption).
For example, two different security administrators may have independently defined the \PIs \etee{c_{a1}}{web_1}{TLS}{c} and \etee{c_{a1}}{s_{c1}}{IPsec}{c, pi}. The former is included in the latter, so that it can be safely removed. In these cases we have a \emph{\redundancy} anomaly (if both the \PIs use the same technology) or an \emph{\inclusion} anomaly (if they use two different protocols).

Another type of suboptimality arises when a tunnel encapsulates other tunnels with a higher security level. This is a \emph{superfluous} anomaly and can be resolved by simply deleting the external, redundant tunnel.

We can also have some channels that can be safely removed without altering the network semantic. This happens in the so called \emph{\internalLoop} anomalies, where a \PI source and destination belong to the same node.

\subsection{\SuboptimalWalks}

A group of \PIs can produce a \emph{\suboptimalWalk} when the path taken by the data is unnecessarily long.

In large networks, a communication between two end-points can take multiple paths, thus generating an \emph{\alternativePath} anomaly.
This is not necessarily a misconfiguration, but nonetheless we detect and report it to the administrators to stay on the safe side.
For example, \etee{g_{c2}}{g_{c3}}{IPsec}{c} forms an \alternativePath w.r.t. \etee{g_{c2}}{g_{c1}}{IPsec}{(c} and \etee{g_{c1}}{g_{c3}}{IPsec}{c}.

Another cause of suboptimality occurs when some data cross a node multiple times during their travel. This \emph{\cyclicPath} anomaly can be removed by deleting the cycles, thus shortening the network path.

\section{\PI hierarchical structure}
\label{sec:Structure}

In this section, we formally define what is a \policyImplementation, its structure, and the relationships between the various network fields that compose it.
In addition, we describe the notion of path used to detect several kinds of network anomalies.
In our model, a \PI $i$ is a tuple:
\begin{equation*}
i = (s, d, t, C, S, G)
\end{equation*}
where:

\begin{itemize}
	\item $s$ and $d$ respectively represent the channel source and destination (Section~\ref{subsec:Entities});
	\item $t$ is the adopted security technology (Section~\ref{subsec:Technologies});
	\item $C$ is an ordered set of coefficients that indicate the required security levels (Section~\ref{subsec:Coefficients});
	\item $S$ is a selector, \ie a tuple of network fields, used to identify the traffic that need to be protected (Section~\ref{subsec:Selectors});
	\item $G$ is the list of the gateways involved in the communication (Section~\ref{subsec:Gateways}).
\end{itemize}

\subsection{Sources ($s$) and destinations ($d$)}
\label{subsec:Entities}

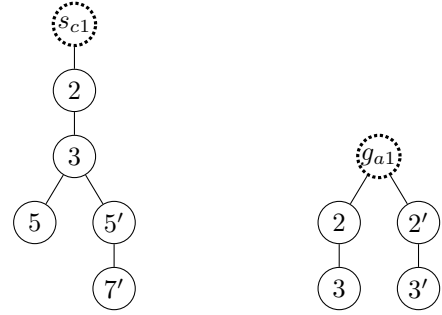
\begin{figure}[ht]
	\centering
	\begin{subfigure}[b]{0.45\linewidth}
		\centering
		\begin{tikzpicture}[level distance=25, sibling distance=30]
	\node[root] {$s_{c1}$}
	child
	{
		node[int] {$2$}
		child
		{
			node[int] {$3$}
			child
			{
				node[int] {$5$}
			}
			child
			{
				node[int] {$5'$}
				child
				{
					node[int] {$7'$}
				}
			}
		}
	};
\end{tikzpicture}
		\caption{Representation of $s_{c1}$.}
		\label{fig:ServerStack}
	\end{subfigure}
	\begin{subfigure}[b]{0.45\linewidth}
		\centering
		\begin{tikzpicture}[level distance=25, sibling distance=30]

		\node[root] (y0) {$g_{a1}$}
			child
			{
				node[int] (y2a) {$2$}
					child
				{
				node[int] (y3a) {$3$}
				}
			}
			child
			{
				node[int] (y2b) {$2'$}
				child
				{
				node[int] (y3b) {$3'$}
				}
			};

\end{tikzpicture}
		\caption{Representation of $g_{a1}$.}
		\label{fig:GatewayStack}
	\end{subfigure}
	\caption{Graphical representation of a server and a gateway.}
	\label{fig:Stack}
\end{figure}
To perform an accurate detection of anomalies, we need to  identify the OSI layer where a communication starts and terminates.
To this purpose, we  use a hierarchical structure that represents the points where the secure communication end-points can be established.
This structure has a very simple tree-like graphical representation as shown in Fig.~\ref{fig:Stack}.

The root represents the network node itself, while all the tree nodes model the available connection end-points, named \emph{network entities}, ordered according to the OSI stack layer (see Fig.~\ref{fig:ServerStack}).
We only focus our attention on the data link, network, session and application layers.
The tree levels may also be associated to layer 2 addresses, IP addresses, port numbers and URIs.
To avoid ambiguity we will use the notation $s_{c1}.l5'$ to specify the node labelled $5'$ in the $s_{c1}$ tree and so on.

Note that the gateways expose multiple interfaces, one for each network where they are connected to.
For instance, in Fig.~\ref{fig:GatewayStack}, the two layer 3 vertices represent the `internal' interface (network $A$) and an `external' interface (the Internet).
If a gateway also supports VPNs via TLS tunnels (\eg OpenVPN), two additional vertices are present at the session level.

Given any two network entities $e_1$ and $e_2$, we define the following relationships:

\begin{itemize}
	\item $e_1$ is \emph{equivalent} to $e_2$ ($e_1 = e_2$) if they are exactly the same entity;
	\item $e_1$ \emph{dominates} $e_2$ ($e_1 \succ e_2$) if all the traffic starting from (or arriving to) $e_2$ passes through $e_1$. On the graph, $e_1$ is an ancestor of $e_2$ in the tree representation. This concept is particularly useful when dealing with security protocols working at different OSI layers. For instance in Fig.~\ref{fig:ServerStack}, $s_{c1}.l3$ dominates $s_{c1}.l7'$;
	\item $e_1$ is a \emph{kin} of $e_2$ ($e_1 \sim e_2$) if $e_1$ and $e_2$ belong to the same network node, but there is no equivalence or dominance relationship amongst them. For example in Fig.~\ref{fig:ServerStack}, $s_{c1}.l5$ is a kin of $s_{c1}.l5'$;
	\item $e_1$ and $e_2$ are \emph{disjoint} ($e_1 \perp e_2$) if they belong to different network nodes (and hence trees).
\end{itemize}

Note that if $e_1$ and $e_2$ are not disjoint ($e_1 \not\perp e_2$) that means that they are on the same device, hence they are related by an equivalence, dominance or kinship relationship.

\subsection{Technologies ($t$)}
\label{subsec:Technologies}

In this paper we take into account a limited set of technologies, but our model is flexible enough to accommodate any security protocol. In particular we will consider only:

\begin{itemize}
	\item for the data link layer: WPA2 and 802.1AE MACsec;
	\item for the network layer: IPsec;
	\item for the session layer: TLS and SSH;
	\item for the application layer: WS-Security.
\end{itemize}

In addition we also use the special NULL technology, indicating that a communication should be created without any kind of protection.

Similar to the network entities, two technologies $t_1$ and $t_2$ can have different relationships:

\begin{itemize}
	\item $t_1$ is \emph{equivalent} to $t_2$ ($t_1 = t_2$), if they are exactly the same technology;
	\item $t_1$ \emph{dominates} $t_2$ ($t_1 \succ t_2$) if $t_1$ operates at an OSI level strictly less than the $t_2$'s one. By definition, the NULL technology is dominated by all the other technologies;
	\item $t_1$ is a \emph{kin} of $t_2$ ($t_1 \sim t_2$) if $t_1$ and $t_2$ are different and work at the same OSI layer;
	\item $t_1$ is \emph{disjoint} from $t_2$ ($t_1 \perp t_2$) if one technology is NULL and the other one is not NULL.
\end{itemize}

In general, the following relationships hold:
\begin{gather*}
t^{(i)} \sim t'^{(i)}, \quad t^{(i)} \ne t'^{(i)}\\
t^{(2)} \succ t^{(3)} \succ t^{(5)} \succ t^{(7)}\\
t \perp \mathrm{NULL}, \quad \forall t \neq \mathrm{NULL}
\end{gather*}

Where $t^{(i)}$ represent a technology at the OSI level $i$.

\subsection{Security coefficients ($C$)}
\label{subsec:Coefficients}

The tuple of  security coefficients consists of several non-negative real values that indicate a required security level for a specific property. The higher a value the stronger the enforcement of a property should be. On the other hand, if a coefficient is zero the related security property must not be enforced. Obviously if the chosen technology is NULL, all the coefficients are zero. These values should be estimated by the administrators with the use of some metrics, for example on the chosen cipher-suite (\eg taking into account the key length, encryption/hash algorithms and cipher mode).

In this paper we focus our attention only on three properties, which are header integrity ($c^{hi}$), payload integrity ($c^{pi}$) and (payload) confidentiality ($c^{c}$), so that:
\begin{equation*}
C =   (c^{hi}, c^{pi}, c^c) 
\end{equation*}
The relationships amongst two coefficient sets $C_1$ and $C_2$ are:

\begin{itemize}
	\item $C_1$ is \emph{equivalent} to $C_2$ ($C_1 = C_2$) if all the coefficients of $C_1$ are the same as their $C_2$'s counterparts;
	\item $C_1$ \emph{dominates} $C_2$ ($C_1 \succ C_2$) if at least one  coefficient of $C_1$ is strictly greater than its $C_2$'s counterparts and the other coefficients of $C_1$ are not less than their $C_2$'s counterparts;
	\item $C_1$ is \emph{disjoint} with $C_2$ ($C_1 \perp C_2$) if there is neither dominance nor equivalence between $C_1$ and $C_2$, that is $C_1 \not\succeq C_2 \wedge C_1 \not\preceq C_2$.
\end{itemize}

\subsection{Selectors ($S$)}
\label{subsec:Selectors}

Some security protocols (\eg IPsec, see RFC-3585) allow the definition of filtering conditions to select the traffic that must be protected.
Our model supports such conditions via the \emph{selectors} $S$ of a policy implementation, that are tuples of network fields.
In theory (and in our model too), $S$ can be arbitrarily defined with any field, however, in practice, the fields in $S$ are usually the well-known five tuple consisting of a source IP address ($ip_{src}$) and port ($p_{src}$), a destination IP address ($ip_{dst}$) and port ($p_{dst}$) and a protocol type ($prt$). We will assume this in the rest of the paper that $S$ at least includes the five-tuple, that is:
\begin{equation*}
S =  (ip_{src}, p_{src}, ip_{dst}, p_{dst}, prt, \ldots )
\end{equation*}

We will use the notation $\overleftarrow{S}$ to indicate a \emph{reverse} list of selectors where source and destination are swapped, that is $\overleftarrow{S} =  (ip_{dst}, p_{dst}, ip_{src}, p_{src}, prt, \ldots )$.
In addition, we use the notation $S|_{f_1 \times f_2 \times \dots}$ to restrict the selector space to the fields $f_1$, $f_2$, $\ldots$.
For instance, in the following sections, we will often use the more compact $S|_{ip_{src} \times p_{src}} =  (ip_{src}, p_{src} )$ instead of the $n$-tuple $ (ip_{src}, *, ip_{dst}, *, *,\dots)$ (where the asterisk symbol $*$ will denote a field matched by any value) to define a selector that matches all the traffic from $ip_{src}$ to $ip_{dst}$ regardless of the port numbers and protocol.
Moreover, the all-matching tuple $ (*, *, *, *, *, \dots) $ will be also shortened to a single $*$ inside a PI definition.

In addition, we will make use of the following relationships between the selector tuples:

\begin{itemize}
	\item $S_1$ is \emph{equivalent} to $S_2$ ($S_1 = S_2$), if their selectors are exactly the same;
	\item $S_1$ \emph{dominates} $S_2$ ($S_1 \succ S_2$) if the matched traffic of $S_2$ is a sub-set of the matched traffic of $S_1$;
	\item $S_1$ is a \emph{kin} to $S_2$ ($S_1 \sim S_2$) if there is at least one communication that matches $S_1$ but not $S_2$ and vice-versa;
	\item $S_1$ is \emph{disjoint} from $S_2$ ($S_1 \perp S_2$) if the sets of the traffic matched by $S_1$ and $S_2$ are disjoint.
\end{itemize}

\subsection{Crossed gateways ($G$)}
\label{subsec:Gateways}
{
	Tunnel \PIs contain an ordered set $G$ that specifies the gateways crossed by the channel traffic. 
	The $G$ sets of tunnel \PIs is statically computed from the network topology and the content of the routing tables.

}
Note that the list of crossed gateways does not contain the channel source and destination nodes.
We use the notation $G^*$ to indicate a list containing also the \PI end-points, that is $G^* = \{ s \} \cup G \cup \{ d \}$. We will also denote the list of crossed gateways in reverse order with $\overleftarrow{G}$.

It is worth presenting an example of \PIs that use gateways.
Given the network in Fig.~\ref{fig:SampleNetwork}, a communication from $c_{a1}$ to $s_{c1}$ that passes into an IPsec tunnel between the two gateways $g_{a1}$ and $g_{c2}$ is implemented by two \PIs:
\begin{align*}
&	i_1 = ( c_{a1}, s_{c1}, \textrm{NULL},  (0, 0, 0) , * , (g_{a1}, g_{c1}, g_{c2})  )\\
&	i_2 = ( g_{a1}, g_{c2}, \textrm{IPsec},  (3, 3, 3) ,(ip_{c_{a1}},*, ip_{s_{c1}},*,*),  (g_{c1})  )
\end{align*}

where $i_1$ specifies the communication that will be encapsulated in the tunnel defined by $i_2$.

\subsection{Paths}

We introduce now the concept of path, which completes the notions used by our model.
The notation $P^{e_1,e_n}$ represents a path starting from the network entity $e_1$ and terminating into the network entity $e_n$. Each \emph{path} is a tuple of \policyImplementations $(i_1, i_2, \dots, i_n)$ where:

\begin{itemize}
	\item the source of the first \PI $i_1$ is $e_1$;
	\item the destination of the last \PI $i_n$ is $e_n$;
	\item given two consecutive \PIs in the path $i_j$ and $i_{j + 1}$, the property $d_j \in S_{j + 1}$ holds.
\end{itemize}

For instance, a path from $c_{c2}$ to $s_{c2}$ is:
\begin{align*}
&	i_1 = ( c_{c2}, g_{c3}, \textrm{NULL},  (0, 0, 0), *, \varnothing )\\
&	i_2 = ( g_{c3}, g_{c2}, \textrm{IPsec},  (3, 3, 3),  (subnet_{c_{c}} ,*,subnet_{c_{s}} ,*,*), \varnothing )\\
&	i_3 = ( g_{c2}, s_{c2}, \textrm{NULL},  (0, 0, 0), *, \varnothing )
\end{align*}
Since $P^{e_1,e_n}$ is a set, we will use the notation $|P^{e_1, e_2} |$ to indicate its cardinality, that is the number of \policyImplementations that compose it.

Note that two paths $P_1^{e_1, e_n}$ and $P_2^{e_1, e_n}$ are different ($P_1^{e_1,e_n} \ne P_2^{e_1,e_n}$) if they differ by at least one element and/or if their respective \PIs are placed in different orders.

\section{Anomaly analysis and resolution}
\label{sec:Analysis}

Having formalized the definition of a \policyImplementation, we can now express the logic formulas used to detect the various anomalies.

\begin{figure}[ht]
	\centering
	\scalebox{0.75}
	{

\begin{tikzpicture}[node distance = 5mm, semithick, draw = gray]
\node (IntraChannel) {\internalLoop};
\node[below = of IntraChannel.west, anchor = west] (Misplaced) {\outOfPlace};
\node[below = of Misplaced.west, anchor = west] (NonEnforceable) {\nonEnforceability};
\node[below = of NonEnforceable.west, anchor = west] (Inadequate) {\inadequacy};
\node[below = of Inadequate.west, anchor = west] (Shadowing) {\shadowing};
\node[below = of Shadowing.west, anchor = west] (Redundancy) {\redundancy};
\node[below = of Redundancy.west, anchor = west] (Exception) {\exception};
\node[below = of Exception.west, anchor = west] (Correlation) {\correlation};
\node[below = of Correlation.west, anchor = west] (Inclusion) {\inclusion};
\node[below = of Inclusion.west, anchor = west] (Affine) {\affinity};
\node[below = of Affine.west, anchor = west] (Hide) {\contradiction};

\node[below = of Hide.west, anchor = west] (Loop) {cycle};
\node[below = of Loop.west, anchor = west] (Monitorable) {\monitorability};
\node[below = of Monitorable.west, anchor = west] (Alternative) {\alternativePath};
\node[below = of Alternative.west, anchor = west] (Superfluous) {\superfluous};
\node[below = of Superfluous.west, anchor = west] (Filtered) {\filteredChannel};
\node[below = of Filtered.west, anchor = west] (Ldue) {\LTwo};
\node[below = of Ldue.west, anchor = west] (Skewed) {\skewedChannel};
\node[below = of Skewed.west, anchor = west] (Asymmetric) {\asymmetricChannel};

\node[fit = (IntraChannel) (Misplaced)] (IrrelevantBox) {};
\node[fit = (NonEnforceable) (Inadequate)] (SecurityBox) {};
\node[fit = (Shadowing) (Redundancy) (Exception) (Correlation)] (IntraBox) {};
\node[fit = (Inclusion) (Affine)  (Hide)  ] (InterBox) {};
\node[fit = (Loop) (Monitorable) (Alternative)] (PathBox) {};
\node[fit = (Superfluous) (Filtered) (Ldue) (Skewed) (Asymmetric) ] (ChannelBox) {};

\node[left = \NodeDistanceC of IrrelevantBox, anchor = west] (Irrelevant) {irrelevant};
\node[left = \NodeDistanceC of SecurityBox, anchor = west] (Security) {\shortstack{unsuitable \\requirements}};
\node[left = \NodeDistanceC of IntraBox, anchor = west] (Intra) {\shortstack{intra\\technology}};
\node[left = \NodeDistanceC of InterBox, anchor = west] (Inter) {\shortstack{inter\\ technology}};
\node[left = \NodeDistanceC of PathBox, anchor = west] (Path) {path};
\node[left = \NodeDistanceC of ChannelBox, anchor = west] (Channel) {channel};

\node[fit = (Irrelevant) (Security)] (PIBox) {};
\node[fit = (Intra) (Inter)] (NodeBox) {};
\node[fit =  (Channel) (Path) ] (NetworkBox) {};

\node[left = \NodeDistanceB of PIBox, anchor = west] (PI) {\shortstack{\PI \\level}};
\node[left = \NodeDistanceB of NodeBox, anchor = west] (Node) {\shortstack{node\\ level}};
\node[left = \NodeDistanceB of NetworkBox, anchor = west] (Network) {\shortstack{network \\level}};

\node[fit = (PI) (Node) (Network)] (AnomaliesBox) {};

\node[left = \NodeDistanceA of Node, anchor = west] (Anomalies) {anomalies};

\REdge{Anomalies}{PI};
\SEdge{Anomalies}{Node};
\REdge{Anomalies}{Network};

\REdge{PI}{Irrelevant};
\REdge{PI}{Security};

\REdge{Node}{Intra};
\REdge{Node}{Inter};

\REdge{Network}{Channel};
\REdge{Network}{Path};

\REdge{Irrelevant}{IntraChannel};
\REdge{Irrelevant}{Misplaced};

\REdge{Security}{NonEnforceable};
\REdge{Security}{Inadequate};

\REdge{Intra}{Shadowing};
\REdge{Intra}{Redundancy};
\REdge{Intra}{Exception};
\REdge{Intra}{Correlation};

\REdge{Inter}{Inclusion};
\SEdge{Inter}{Affine};
\REdge{Inter}{Hide};

\REdge{Channel}{Superfluous};
\REdge{Channel}{Filtered};
\SEdge{Channel}{Ldue};
\REdge{Channel}{Skewed};
\REdge{Channel}{Asymmetric};

\REdge{Path}{Loop};
\SEdge{Path}{Monitorable};
\REdge{Path}{Alternative};

\end{tikzpicture}

	}
	\caption{Information-centric taxonomy of anomalies.}
	\label{fig:TechnicalTaxonomy}
\end{figure}
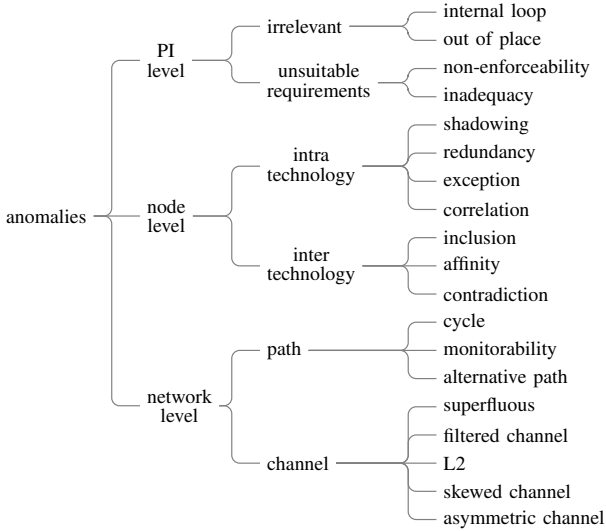

In Section~\ref{sec:Example}, we introduced an anomaly classification based on five macro-categories (Fig.~\ref{fig:MacroCategoriesTaxonomy}), which emphasizes the side effects of an anomaly. However in the following paragraphs,  we will use a more technical classification (Fig.~\ref{fig:TechnicalTaxonomy}), better suited for a more formal discussion. In fact, such classification highlights the possible levels of interactions among \PIs and, hence, at which level the anomaly is generated. We distinguish three levels of anomalies:
\begin{enumerate*}
	\item the \emph{\PI level anomalies} that occur within a single \PI;
	\item the \emph{node level anomalies}, which come up between two distinct \PIs placed on the same node;
	\item the \emph{network level anomalies } arising between distinct \PIs that belong to different nodes.
\end{enumerate*}
\subsection{\PI level anomalies}

We can distinguish two families of \PI level anomalies: irrelevant and unsuitable requirement anomalies. A \PI that generates an \emph{irrelevant anomaly} (which splits into \internalLoop and \outOfPlace) is meaningless for the network semantics, so that their presence does not change how a network exchanges the data. The \emph{unsuitable requirement anomalies} (\nonEnforceability and \inadequacy) instead break some security requisite and they can lead to severe problems.

\subsubsection{\InternalLoop~--~$\mathcal{A}_{il}(i_1)$}

There is an \emph{\internalLoop anomaly} when the source and destination end-points are on the same node, thus creating a communication loop.
These anomalies can be inferred by using the formula:
\begin{equation}
\mathcal{A}_{il}(i_1) \Leftrightarrow s_1 \not\perp d_1
\end{equation}

The proposed resolution method is to simply delete $i_1$.

\subsubsection{\OutOfPlace~--~$\mathcal{A}_{op}(i_1)$}

There is an \emph{\outOfPlace anomaly} when a \PI is deployed on a wrong network node. That means that the source is disjoint with the node where the \PI is deployed. To detect these anomalies, we use the function $\mathcal{N}(i_1)$ that returns the node where the \PI is actually deployed. The formula is then:
\begin{equation}
\mathcal{A}_{op}(i_1) \Leftrightarrow \mathcal{N}(i_1) \perp s_1
\end{equation}
The simplest resolution is to delete $i_1$. However, a more suitable approach can be to redeploy the \PI on the correct node or to appropriately modify its source.

\subsubsection{\NonEnforceability~--~$\mathcal{A}_{ne}(i_1)$}

A \PI $i_1$  is non-enforceable when its technology is not supported by  the source, the destination or when its security coefficients are `too high', and hence cannot be enforced (e.g., due to missing strong encryption algorithms).

We will make use of two functions: $\mathcal{T}(e)$, which returns the set of technologies supported by the node $e$, and $C_{max}(i_1)$, which returns the set of maximum enforceable coefficients by the \PI $i_1$.
These anomalies can be identified with the formula:
\begin{equation*}
\mathcal{A}_{ne}(i_1) \Leftrightarrow C_1 \succ \mathcal{C}_{max}(i_1) \vee t_1\not\in \mathcal{T}(s_1) \vee t_1\not\in \mathcal{T}(d_1)
\end{equation*}
To resolve these anomalies, an administrator can choose to upgrade the security libraries/services on the \PI source/destination to support the desired technologies or, alternatively, he might modify the \PI by changing the protocol or lowering the security coefficients (at his own risk).

\subsubsection{\Inadequacy~--~$\mathcal{A}_{in}(i_1)$}

We have an \emph{\inadequacy anomaly} when the security coefficients of a \policyImplementation establish a channel with a security that is lower than an acceptable threshold. We can use a function $C_{min}(i_1)$ that returns the minimum acceptable coefficients for the channel defined by the \PI $i_1$.
This function should be defined a priori by the administrators according to a (corporate) metric or best practice~\cite{NSA,NIST,ISO}.
For example a network administrator could define a function such as:
\begin{equation*}
C_{min}(i_1) =
\begin{cases}
(1, 1, 1)  & \text{if $i_1$ is crossing the Internet}\\
(0, 0, 0)  & \text{otherwise}\\
\end{cases}
\end{equation*}
We can detect these anomalies with the rule:
\begin{equation}
\mathcal{A}_{in}(i_1) \Leftrightarrow C_1 \prec \mathcal{C}_{min}(i_1)
\end{equation}
In order to fix these issues, the security requirements of the \policyImplementation must be increased so that the property $C_1 \succeq \mathcal{C}_{min}(i_1)$ holds.

\subsection{Node level anomalies}

A \emph{node level anomaly} occurs between two distinct \policyImplementations laying on the same node.

If the two \PIs have the same technology then we have an \emph{intra-technology anomaly} (\shadowing, \exception, \redundancy and \correlation), otherwise we have an \emph{inter-technology anomaly} (\inclusion, \affinity and \contradiction). The intra-technology anomaly category has been heavily inspired by the work of Hamed\etal's~\cite{AlIPsec}.

For detecting these anomalies, we assume that the two \PIs have the same crossed gateways, that is $G_1 = G_2$. In addition, we will also make use of the function $\pi(i) \in \mathbb{N}$ that returns the priority of a \PI in a \PIset (the lower the number the higher the priority).

\subsubsection{\Shadowing~--~$\mathcal{A}_{sh}(i_1, i_2)$}

A \PI $i_2$ is \emph{shadowed} when there is another \policyImplementation $i_1$ with a higher priority that matches all the traffic of the first one
($s_1 \succeq s_2 \wedge d_1 \succeq d_2 \wedge S_1 \succeq S_2$) and has disjoint  security coefficients. We can detect these anomalies using the formula:
\begin{gather}
\mathcal{A}_{sh}(i_1, i_2) \Leftrightarrow \pi(i_1) < \pi(i_2) \wedge t_1 = t_2 \wedge s_1 \succeq s_2 \wedge \nonumber \\
d_1 \succeq d_2 \wedge S_1 \succeq S_2  \wedge C_1 \perp C_2 \wedge G_1 = G_2 \wedge i_1 \neq i_2
\end{gather}

In order to resolve these kind of anomalies, either the shadowed \PI is deleted or the two \PIs are replaced by another \PI $i_3$ that is an upper bound of the previous ones. In particular, $i_3$ will have the following fields:

\begin{itemize}
	\item $s_3$ is the least upper bound of $s_1$ and $s_2$ such that $s_3 \succeq s_1$ and $s_3 \succeq s_2$ hold;
	\item $d_3$ is the least upper bound of $d_1$ and $d_2$ such that $d_3 \succeq d_1$ and $d_3 \succeq d_2$ hold;
	\item $C_3 = \{ c_{3, i} \}_i$ can be computed as $c_{3, i} = \max(c_{1, i}, c_{2, i})$ where $C_1 = \{ c_{1, i} \}_i$ and $C_2 = \{ c_{2, i} \}_i$;
	\item  $S_3$ is the least upper bound of $S_1$ and $S_2$ such that $S_3 \succeq S_1$ and $S_3 \succeq S_2$ hold;
	\item $t_3 = t_1 = t_2$,  $G_3 = G_1 = G_2$.
\end{itemize}

To maintain the semantics of the system, the new \PI $i_3$ should be inserted at the highest priority (\ie $\pi(i_1)$).

\subsubsection{\Redundancy~--~$\mathcal{A}_{re}(i_1, i_2)$}

A \PI $i_2$ is \emph{redundant} when there is another \policyImplementation $i_1$ with a higher priority that matches all the traffic of the first one and its security coefficients are equal or dominates the other \PI's coefficients. The following formula can be used to infer these problems:
\begin{gather}
\mathcal{A}_{re}(i_1, i_2) \Leftrightarrow t_1 = t_2 \wedge s_1 \succeq s_2 \wedge  \nonumber\\
d_1 \succeq d_2 \wedge S_1 \succeq S_2 \wedge C_1 \succeq C_2  \wedge G_1 = G_2 \wedge i_1 \neq i_2
\end{gather}

The proposed resolution is to delete $i_2$, because it does not add new semantics to the policy.

\subsubsection{\Exception~--~$\mathcal{A}_{ex}(i_1, i_2)$}

A \PI $i_2$ is an \emph{exception} of another \policyImplementation $i_1$ with a higher priority if they have disjoint security coefficients and $i_2$ is a superset match of $i_1$ ($  s_1 \prec s_2 \wedge d_1 \prec d_2  \wedge S_1 \prec S_2 $) . The relative detection formula is:
\begin{gather}
\mathcal{A}_{ex}(i_1, i_2) \Leftrightarrow \pi(i_1) \prec \pi(i_2) \wedge t_1 = t_2 \wedge s_1 \prec s_2  \wedge \nonumber\\
d_1 \prec d_2 \wedge S_1 \prec S_2 \wedge C_1 \perp C_2 \wedge G_1 = G_2 \wedge i_1 \neq i_2
\end{gather}

Exceptions are analogous to the \shadowing anomalies (just the opposite order of precedences) thus they share the same resolution approach.

\subsubsection{\Correlation~--~$\mathcal{A}_{co}(i_1, i_2)$}

A \PI $i_2$ is \emph{correlated} with another \policyImplementation $i_1$ if they have disjoint security coefficients, $i_1$ matches some traffic for $i_2$ and vice versa. In other words, the source and destination of  $i_1$ and $i_2$ belong to the same node and there is no other intra-technology anomaly between \policyImplementations  (\ie shadowing, redundancy or exception). We can detect these anomalies via the formula:
\begin{gather}
\mathcal{A}_{co}(i_1, i_2) \Leftrightarrow s_1 \not\perp s_2 \wedge d_1 \not\perp d_2 \wedge t_1 = t_2 \nonumber  \wedge\\
S_1 \not\perp S_2 \wedge G_1 = G_2  \wedge 	 \neg\mathcal{A}_{sh}(i_1, i_2)  \nonumber \wedge  \\
\neg\mathcal{A}_{ex}(i_1, i_2) \wedge \neg\mathcal{A}_{re}(i_1, i_2) \wedge i_1 \neq i_2
\end{gather}

To resolve these anomalies, the two \PIs $i_1$ and $i_2$ can be replaced with a new \PI $i_3$ with the same fields as described in the \shadowing anomaly resolution technique. However, the newly created \policyImplementation will be inserted with a priority $\pi(i_3) = \min(\pi(i_1), \pi( i_2))$.

\subsubsection{\Inclusion~--~$\mathcal{A}_{in}(i_1, i_2)$}

The \PI $i_1$ \emph{includes} (or dominates) the \policyImplementation $i_2$ when all fields of $i_1$ dominate or are equal to their respective $i_2$ fields, except one that is strictly dominant. We can detect these anomalies with:
\begin{gather} 
\mathcal{A}_{in}(i_1, i_2) \Leftrightarrow s_1 \succeq s_2 \wedge d_1 \succeq d_2 \wedge t_1 \succeq t_2 \wedge  \nonumber\\
C_1 \succeq C_2 \wedge S_1 \succeq S_2 \wedge G_1 = G_2 \wedge i_1 \neq i_2
\end{gather}
The simplest way to resolve these anomalies is to delete $i_2$ (the `innermost' \PI). However, an administrator can also keep both the \policyImplementations for a security-in-depth approach.

\subsubsection{\Affinity~--~$\mathcal{A}_{a\!f}(i_1, i_2)$}

A \PI $i_1$ is \emph{affine} with another \policyImplementation $i_2$ when they share some fields, but none of the \PIs includes the other. We can detect these anomalies with the formula:
\begin{gather}
\mathcal{A}_{af}(i_1, i_2) \Leftrightarrow s_1 \not\perp s_2 \wedge d_1 \not\perp d_2 \wedge t_1 \not\perp t_2 \wedge  \nonumber\\
S_1 \not\perp S_2 \wedge \neg\mathcal{A}_{in}(i_1, i_2) \wedge \neg\mathcal{A}_{in}(i_2, i_1) \wedge i_1 \neq i_2
\end{gather}

To resolve this type of anomalies, the two \PIs should be replaced with a new \PI $i_3$ that is an upper bound of the previous ones:

\begin{itemize}
	\item $s_3$ is the least upper bound of $s_1$ and $s_2$ such that $s_3 \succeq s_1$ and $s_3 \succeq s_2$ hold;
	\item $d_3$ is the least upper bound of $d_1$ and $d_2$ such that $d_3 \succeq d_1$ and $d_3 \succeq d_2$ hold;
	\item $t_3$ is the least upper bound of $t_1$ and $t_2$ such that $t_3 \succeq t_1$ and $t_3 \succeq t_2$ hold;
	\item $C_3 = \{ c_{3, i} \}_i$ can be computed as $c_{3, i} = \max(c_{1, i}, c_{2, i})$ where $C_1 = \{ c_{1, i} \}_i$ and $C_2 = \{ c_{2, i} \}_i$;
	\item $S_3$ is the least upper bound of $S_1$ and $S_2$ such that $S_3 \succeq S_1$ and $S_3 \succeq S_2$ hold;
	\item $G_3 = G_1 = G_2$ .
\end{itemize}

\subsubsection{\Contradiction~--~$\mathcal{A}_{co}(i_1, i_2)$}

Two \PIs $i_1$ and $i_2$ are in a \emph{contradiction} if their sources/destinations lay on the same node but their technologies are disjoint (that is one \PI is using the NULL technology and the other one a security protocol). The formula for detecting these anomalies is:
\begin{gather}
\mathcal{A}_{co}(i_1, i_2) \Leftrightarrow s_1 \not\perp s_2 \wedge d_1 \not\perp d_2 \wedge \nonumber\\
t_1 \perp t_2 \wedge  S_1 \not\perp S_2 \wedge i_1 \neq i_2
\end{gather}

The resolution is the removal of one \PI, however, due to the high ambiguity of the situation is up to the administrator to choose which one (we cannot automatically pick among `protect' and `do not protect').

\subsection{Network level anomaly}

The network level anomalies occur between distinct \PIs that belong to different nodes. We can split these anomalies into two main categories: path (\cyclicPath, \monitorability and \alternativePath anomalies), and channel anomalies (\superfluous, \filteredChannel, \LTwo, \skewedChannel and \asymmetricChannel anomalies).

\subsubsection{\Superfluous~--~$\mathcal{A}_{su}(i_1)$}

A \PI $i_1$ is \emph{\superfluous} if it models a tunnel that protects less than all its inner end-to-end channels. That is, the security coefficient of a superfluous tunnel $i_1$ are smaller than all the encapsulated channels ($\forall \ i_k : s_k \in S_1|_{ip_{src}\times p_{src}\times\dots} \wedge G_k^* \supset G_1^*$). 
This anomalous \PIs can be detected by using the formula:
\begin{gather}
\mathcal{A}_{su}(i_1) \Leftrightarrow  \nexists \ i_k  : s_k \in S_1|_{ip_{src}\times p_{src}\times\dots} \wedge \nonumber\\
G_k^* \supset G_1^* \wedge C_k \prec C_1
\end{gather}

Since all the data transported in the tunnel are better protected than the tunnel itself, the obvious resolution is to delete $i_1$ (since it is \superfluous). However, as in the \inclusion anomaly, an administrator could choose to keep the \PI to (slightly) increase the security of the network.

\subsubsection{\SkewedChannel~--~$\mathcal{A}_{sk}(i_1,i_2)$}

Two \PIs $i_1$ and $i_2$ that define two tunnels are \emph{skewed} if their respective channels overlap. This type of anomalies are tricky because in a portion of the network the traffic will be sent without any form of encryption (Fig.~\ref{fig:TrellisDiagram}).  We can detect these anomalies with:
\begin{gather}
\mathcal{A}_{sk}(i_1,i_2) \Leftrightarrow s_1 \in S_2|_{ip_{src} \times p_{src}\times\dots} \wedge (\vert G_1^* \cap G_2^* \vert) \geqslant 2 \wedge \nonumber\\
(G_2^* \setminus G_1^* \neq 0) \wedge c_1^c >0   \wedge c_2^c >0 \wedge i_1 \neq i_2
\end{gather}

In order to resolve this kind of anomalies, the two \PIs must be split in three (or more) non-overlapping \policyImplementations.

\subsubsection{\FilteredChannel~--~$\mathcal{A}_{fi}(i_1)$}

A \PI $i_1$ is \emph{filtered} when there exists at least one node $e$ in its path with a filtering rule that discards all its traffic. Given a function $\mathcal{F}_e(i_1)$, which returns true if the traffic related to $i_1$ is dropped and false otherwise, we can formally model this anomaly with:
\begin{equation}
\mathcal{A}_{fi}(i_1)\Leftrightarrow \exists e : e\in G_1 \wedge \mathcal{F}_e(i_1) = true
\end{equation}
In practice, the output of the function $\mathcal{F}_e(i_1)$ can be populated either by means of a network reachability analysis~\cite{Basile2013} or by using some firewall policy queries~\cite{Khakpour2013}.

This anomaly is particularly severe since it completely hinders the connectivity between a number of network nodes. To remove the problem, the administrator can choose to delete the \PI $i_1$ or modify accordingly the filtering rule.

\subsubsection{\LTwo~--~$\mathcal{A}_{L2}(i_1)$}

An \emph{\LTwo anomaly} occurs when a \PI that uses a data-link layer technology  crosses an area using a different layer 2 protocol. For instance, we have a \LTwo anomaly when a WPA2 \policyImplementation crosses some Ethernet nodes, so that we cannot use WPA2 for the whole path. We can express this anomaly by using a function $\mathcal{T}^{(2)}(e)$ that returns the set of technologies at layer 2 supported by the node $e$. We can then write the formula:
\begin{equation}
\mathcal{A}_{L2}(i_1)\Leftrightarrow \exists e: e\in G_1^* \wedge t_1 \not\in \mathcal{T}^{(2)}(e)
\end{equation}
These anomalies are quite hard to resolve since they require a complete edit of the \PI, by choosing a technology at a layer strictly greater than 2.

\subsubsection{\AsymmetricChannel~--~$\mathcal{A}_{as}(i_1)$}

A \PI $i_1$ is \emph{asymmetric} if does not exist another \PI with:
\begin{enumerate*}
	\item the source and destination swapped ($s_1 \not\perp d_2 \wedge d_1 \not\perp s_2$);
	\item the same technology and security coefficients;
	\item the same list of crossed gateways, but in reverse order.
\end{enumerate*}
In other words, these problems arise when we have a bidirectional communication with a channel weaker than the other. We can identify these anomalies by using the formula:
\begin{gather}
\mathcal{A}_{as}(i_1) \Leftrightarrow \nexists i_2 : s_1 \not\perp d_2 \wedge d_1 \not\perp s_2 \wedge  \nonumber\\
t_1 = t_2 \wedge C_1 = C_2 \wedge S_1 \not\perp \overleftarrow{S_2 }\wedge G_1 = \overleftarrow{G_2}
\end{gather}

Administrators must check these anomalies to verify if they represent the wanted protection.

\subsubsection{\CyclicPath~--~$\mathcal{A}_{cy}(P^{e_1, e_2})$}

There is a \emph{\cyclicPath} anomaly between two nodes $e_1$ and $e_2$ if there is at least one cycle in the path connecting them. These anomalies can be detected with any of the several very efficient algorithms available in literature to perform cycle detection \cite{Tarjan72}.

The only way to resolve this kind of anomalies is to modify the \PIs to remove the cycles.

\subsubsection{\Monitorability~--~$\mathcal{A}_{mo}(P^{e_1, e_2})$}

A path $P^{e_1, e_2}$ is \emph{monitorable} when there is not an end-to-end channel between $e_1$ and $e_2$. That means that, even if the connections are protected by encryption, there is at least one node where an encrypt/decrypt operation is performed, thus potentially breaking the confidentiality of the communication. These anomalies can be detected by using the formula:
\begin{equation}
\mathcal{A}_{mo}(P^{e_1, e_2}) \Leftrightarrow \nexists P^{e_1, e_2} : (|P^{e_1, e_2}| = 1 \wedge  i_j \in P^{e_1, e_2} : c^c_j > 0)
\end{equation}

If the network is not trusted, the obvious way to remove this anomaly is to edit the \PIs such that there are only end-to-end channels between $e_1$ and $e_2$ .

\subsubsection{\AlternativePath~--~$\mathcal{A}_{al}(P^{e_1, e_2}_1, P^{e_1, e_2}_2)$}

There is an \emph{\alternativePath} between two nodes $e_1$ and $e_2$ if there are two or more different paths that can be taken from the source node to the destination node.
These anomalies can be easily found by using the formula:
\begin{equation}
\mathcal{A}_{al}(P^{e_1, e_2}_1, P^{e_1, e_2}_2) \Leftrightarrow \exists P_1^{e_1, e_2}, P_2^{e_1, e_2} : P^{e_1, e_2}_1\neq  P^{e_1, e_2}_2
\end{equation}

To remove this redundancy, the administrators have to choose the `best' path for the communication and delete the other ones.
The choice can be made by using different strategies, such as picking the shortest path or the path containing the \PIs with the highest security coefficients.


\section{Graph-based representation of the anomalies}
\label{sec:Graph}

Aiming for a model that also has practical relevance, we investigated the possibility of a user friendly representation of our anomalies, as logical formulas are not easily usable by administrators.
In Section~\ref{sec:Structure} we  have already sketched our hierarchical view of a network node.
This allows the representation of secure communications by connecting network nodes to form a multi-graph.
The obvious advantage of such representation is that it allows a network administrator to visualize the communications at a glance. Our claim is that it allows the intuitive identification of the anomalies and the consequences and the reactions.

\begin{figure}[t]
	\centering
	\begin{tikzpicture}[level distance=25, sibling distance=30]
			\node[root](x0) {$c_{a1}$}
			child
			{ 
				node[int](x2b){$2$}
				child
				{
					node[int] (x3b) {$3$}
					child
					{
						node[int] (x5b) {$5$}
						child
						{
							node[int] (x7b) {$7$}
						}
					}
				}
			};

		\node[root,right of=x0,right=2.5] (y0) {$s_{c1}$}
		child
		{
			node[int] (y2a) {$2$}
			child
			{
				node[int](y3a){$3$}
				child
				{
					node[int] (y5a) {$5$}
				}
				child
				{
					node[int] (y5b) {$5'$}
					child
					{
						node[int] (y7a) {$7'$}
					}
				}
			}
		};

\path[ione, bend left = 20] (x3b) edge node[coef] {$(*, *, *, *, *)$} node[tech] {IPsec: $(0, 0,3)$} (y3a);
\path[itwo, bend right = 20] (x5b) edge node[coef] {$(*, *, *, *, *)$} node[tech] {TLS: $(0, 3, 0)$} (y5a);
\end{tikzpicture}
	\caption{Graphical representation of an \affinity anomaly.}
	\label{fig:AffinityGraph}
\end{figure}
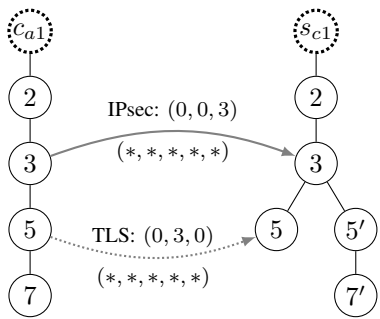

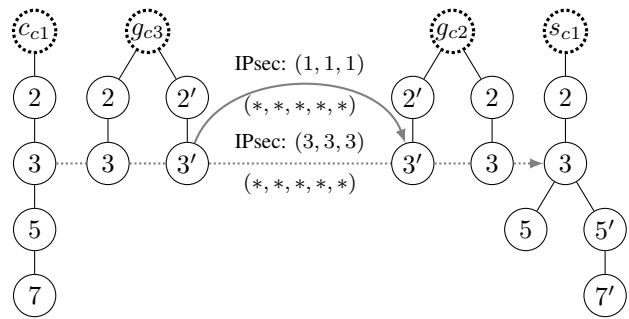
\begin{figure}[t]
	\centering
	\begin{tikzpicture}[level distance=25, sibling distance=30]
		\node[root] (x0) {$c_{c1}$}
			child
			{
				node[int] (x2a) {$2$}
				child
				{
					node[int] (x3a) {$3$}
					child
					{
						node[int] (x5a) {$5$}
						child
						{
							node[int] (x7a) {$7$}
						}
					}
				}
			};

		\node[root,right of=x0,right=0.2] (y0) {$g_{c3}$}
			child
			{
				node[int] (y2a) {$2$}
					child
				{
				node[int] (y3a) {$3$}
				}
			}
			child
			{
				node[int] (y2b) {$2'$}
				child
				{
				node[int] (y3b) {$3'$}
				}
			};

		\node[root,right of=y0,right=2.75] (z0) {$g_{c2}$}
			child
			{
			  node[int] (z2a) {$2'$}
			  	child
			  	{
					node[int] (z3a) {$3'$}
				}
			}
			child
			{
				 node[int] (z2b) {$2$}
				 child
				 {
				node[int] (z3b) {$3$}
				}
			};

		\node[root,right of=z0,right=0.2] (w0) {$s_{c1}$}
		child
		{
			node[int] (w2a) {$2$}
			child
			{
				node[int] (w3a) {$3$}
				child
				{
					node[int] (w5a) {$5$}
				}
				child
				{
					node[int] (w5b) {$5'$}
					child
					{
						node[int] (w7a) {$7'$}
					}
				}
			}
		};

	\path[ione, bend left = 70] (y3b) edge node[coef] {$(*, *, *, *, *)$} node[tech] {IPsec: $(1,1, 1)$} (z3a);
	\path[itwo, -] (x3a) edge (y3a);
	\path[itwo, -] (y3a) edge (y3b);
	\path[itwo, -] (y3b) edge node[coef] {$(*, *, *, *, *)$} node[tech] {IPsec: $(3, 3, 3)$} (z3a);
	\path[itwo, -] (z3a) edge (z3b);
	\path[itwo] (z3b) edge (w3a);
\end{tikzpicture}
	\caption{Graphical representation of a \superfluous anomaly.}
	\label{fig:SuperfluousGraph}
\end{figure}

For example, Fig.~\ref{fig:AffinityGraph} shows an \affinity anomaly between two \PIs. The first \policyImplementation (solid line) enforces IPsec in transport mode and requires only confidentiality, while the second one (dashed line) uses TLS and enforces only payload integrity. The graph clearly shows that the two \PIs are correlated as there are two ``parallel'' arrows. These \PIs are affine since they share some common aspects, but none of them includes the other one. An administrator may understand that he can alternatively use (1) only the IPsec channel, if he adds also payload and header integrity, (2) only the TLS channel, if he adds confidentiality, or (3) keep both.
Another example is shown in Fig.~\ref{fig:SuperfluousGraph}, where a \superfluous anomaly is depicted. We recall that a channel is \superfluous when there is another tunnel that covers at least the same traffic but protects the communication with a higher security level.
In this case the IPsec tunnel between $g_{c3}$ and $g_{c2}$ is redundant, so an administrator immediately sees that it can be safely removed.

All the anomalies (but the \outOfPlace one) have a corresponding graphical representation.
These representations can be built by including the network node trees corresponding to the communication end-points (Fig.~\ref{fig:Stack}), that is, the source and the destination of the \PI.
The  \PIs that enforce end-to-end channels are represented as single directed edges between two communication vertices, \ie the proper communication layer nodes. For instance, in Fig.~\ref{fig:AffinityGraph} the edge connects the layer 3 nodes as the technology is IPsec.
To increase the expressiveness of our representation, each edge is also labelled with the technology, the security coefficients required by the \PI and the selectors.
%
To represent the \policyImplementations that enforce site-to-site and remote-access communications, we add all the network node trees corresponding to the crossed gateways and an edge crossing all the communication parties.
For instance, in case of a tunnel (Fig.~\ref{fig:SuperfluousGraph}), we introduce an edge crossing the source node, the first gateway, the second gateway and terminating into the destination node.

As anticipated, \outOfPlace is the only anomaly that we do not represent graphically.
{
	Visualizing this anomaly negligibly boosts practical usefulness of our graphical representation but significantly increases its complexity.}


%

\begin{table*}[t]
	\centering
	\scalebox{0.9}{
		\begin{tabular}{lccccc}
			\toprule
			expertise & \insecureCommunications & \unfeasibleCommunications & \potentialErrors & \suboptimalImplementations & at least one type\\
			\cmidrule{1-6}
			low & 70.00\% & 60.00\% & 60.00\%  & 70.00\% & 100.00\%  \\		
			medium & 60.00\% & 30.00\% & 50.00\%  & 40.00\% & 90.00\% \\		
			high & 30.00\% & 20.00\% & 20.00\%  & 70.00\% & 90.00\%\\
			\cmidrule{1-6}
			average & 53.33\% & 36.67\% & 43.33\%  & 60\% & 93.33\%\\
			\bottomrule
		\end{tabular}
	}
	\caption{Percentage of administrators that created at least one anomaly in a macro-category.}
	\label{Tab1b}
\end{table*}

\begin{table*}[t]
	\centering
	\scalebox{0.9}{
		\begin{tabular}{lccccccccc}
			\toprule		
			expertise & \internalLoop & \nonEnforceability & \inadequacy & \inclusion  & \affinity & \monitorability & \superfluous & \filtered &  \contradiction\\
			\cmidrule{1-10}
			low & 20.00\% & 30.00\% & 40.00\% &  30.00\% & 50.00\%  & 30.00\% & 30.00\%  & 30.00\% & 30.00\%\\
			medium& 10.00\% & 20.00\% & 40.00\% &  20.00\% & 40.00\%  & 20.00\% & 30.00\%  & 10.00\% & 10.00\% \\
			high & 10.00\% & 10.00\% & 10.00\% &  20.00\% & 20.00\%  & 20.00\% & 50.00\%  & 10.00\% & 0.00\% \\		
			\cmidrule{1-10}
			average & 13.33\% & 20.00\% & 30.00\% &   23.33\% & 36.67\%  & 33.33\% & 30.00\%  & 16.33\% & 13.33\% \\		
			\bottomrule
		\end{tabular}
	}
	\caption{Percentage of administrators that created at least one anomaly.}
	\label{Tab1a}
\end{table*}

\begin{table*}[t]
	\centering
	\scalebox{0.9}{
		\begin{tabular}{lccccc}
			\toprule
			expertise & \insecureCommunications & \unfeasibleCommunications & \potentialErrors & \suboptimalImplementations & total\\
			\cmidrule{1-6}
			low &  18.41\% & 22.39\% &  15.92\%&  7.46\% &  64.18\% \\		
			medium &  16.54\% & 12.78\% & 9.77\%&  9.77\% &  48.87\% \\		
			high &  12.90\% & 4.03\% & 2.42\%&  12.90\% & 32.26\%  \\
			\cmidrule{1-6}
			average &  16.38\% & 14.63\% & 10.48\%&  9.61\% & 51.09\%  \\
			\bottomrule
			
		\end{tabular}
	}
	\caption{Percentages of anomalies introduced by the administrators grouped in macro-categories.}
	\label{Tab2a}
\end{table*}

\begin{table*}[t]
	\centering
	\scalebox{0.9}{
		\begin{tabular}{lccccccccc}
			\toprule
			expertise & \internalLoop & \nonEnforceability & \inadequacy & \inclusion & \affinity & \monitorability & \superfluous & \filtered & \contradiction\\
			\cmidrule{1-10}
			low & 1.49\% & 4.48\%  & 10.95\% & 2.99\%  &6.97\% & 2.99\% & 7.46\% & 17.91\% & 8.96\%  \\		
			medium & 1.50\% & 3.76\%  & 13.53\% & 4.51\%  & 5.26\% & 3.01\% & 3.76\% & 9.02\% & 4.51\%  \\		
			high & 1.61\% & 0.81\%  & 4.03\% & 3.23\%  &2.24\% & 8.87\% & 8.06\% & 3.23\% & 0.00\%   \\
			\cmidrule{2-10}
			average & 1.53\% & 3.28\% & 9.83\% & 3.49\% & 5.24\% & 6.55\% & 4.59\% & 11.35\% & 5.24\%  \\
			\bottomrule
		\end{tabular}
	}
	\caption{Percentages of anomalies introduced by the administrators.}
	\label{Tab2b}
\end{table*}


\section{Model validation}
\label{sec:Validation}

In this section, we present the evaluation of our anomaly analysis model's usefulness and usability.

\subsection{Empirical assessment}
\label{sec:Empirical}

In order to evaluate the practical importance of our work, we conducted an empirical assessment. We tried to answer two simple yet interesting research questions:

\begin{enumerate}[label = {RQ\arabic*.}, leftmargin = 2.55em]
	\item are anomalies presented in this paper actually introduced by the administrators when configuring the \cpp?
	\item does the number of anomalies decrease when the administrator expertise grows?
\end{enumerate}

If RQ1 is confirmed, we could deduce that performing the detection can help improving the policy enforcement correctness in real world networks.
RQ2 instead can give us insights on users that can benefit the most from our anomaly analysis model.

We mainly focused on the new kinds of anomalies presented for the first time in this paper. For this reason we did not report statistics on anomalies already present in the literature, namely the \shadowing, \redundancy, \exception, \correlation, the \skewedChannel (overlapping sessions) and \outOfPlace (irrelevances) whose importance was already proved in other original works~\cite{AlIPsec, AlTax,Alfaro08}. We designed the experiment to be completed by the administrators in one hour, therefore we avoided to provide data link information and kept the size of the network reasonably low. For this reason, the L2, \asymmetricChannel, cycle and alternative path anomalies were not considered in our study.

In order to answer the research questions, we conducted an experiment by recruiting a set of 30 administrators.
We split them into three categories according to their expertise level (high, medium and low), each one containing 10 people.
In the test, we have considered as high expertise administrators people with more than two years of experience in the security field, as medium expertise administrators people with more than two years of practice in the (non-security) network field, and as low expertise administrators the remaining ones.

We asked them to enforce five \cpp (\eg ``all the administrators must securely reach the accounting service'') by implementing them as a set of \PIs. The landscape was a small network (consisting of 5 subnets, 6 servers, 9 clients and 10 gateways). 
The network description and the \cpp were available online to the participants both as a web page and as a PDF document to be accessed offline.
The participants were asked to write all \PIs where all the their fields where constrained to valid values (\eg correct node and protocol names) to avoid uninteresting errors. We did not impose neither a time limit nor a maximum number of \PIs.

The analysis of experiments data gave us very interesting information.
First of all, 93\% of administrators introduced at least one anomaly, regardless of the expertise, as shown in Table~\ref{Tab1b}. In addition, all the new anomalies have been introduced by at least one administrator (Table~\ref{Tab1a}). Interestingly enough, all the anomaly types except contradictions were also introduced by expert administrators. This result successfully answered positively the research question RQ1, that is the anomalies presented in this paper can appear in real world scenario, hence it is useful to look for them.

The RQ2 research question (the more the expertise of administrators the less the anomalies) has been also confirmed for all the anomaly macro-categories except one, the \suboptimalImplementations (Table~\ref{Tab2a}). Obviously having a better understanding of a network and its different security controls, reduces the chance of introducing anomalies. This is particularly evident for the \filtered anomalies, as the administrators also have to consider the interactions with firewalls to avoid them, but it is also valid for the \nonEnforceability, \inadequacy, \affinity, and \contradiction anomalies (Table~\ref{Tab2b}). On the other hand, the \suboptimalImplementations tend to increases because the expert administrators add more \superfluous anomalies, most likely for providing a defense in depth approach, although this was not expressly required in the exercise. Moreover, expert administrators' \PIs also contain several \monitorability anomalies, since they tend to make an extensive use of tunnels while the less skilled ones mainly use end-to-end channels. In short, the experienced administrators tend to break secure communications to improve the overall network performance. In this sample network, the \monitorability anomalies are not the most serious issues (as we had homogeneous security levels in all networks), however, in general it is worth checking them.
Finally, there is a number of \internalLoop anomalies, constant with the expertise, probably due to distraction errors.

The statistical significance of the differences between the results on the three expertise levels, has been assessed with the analysis of variances (ANOVA) with a significance level of 0.05. We performed the test on the number of error types and errors for each administrator and obtained two P-values of 0.034 and 0.006, thus successfully proving the hypothesis.


We identified two major threats to the validity of our experiment.
First, for practical reasons (anonymity, expected number of participants, web-based data collection) we provided a single administration task, therefore, the experiment is subject to the mono-operational bias. 
This threat has been mitigated as the proposed task was realistic and complex enough to capture the heterogeneity of administrators' tasks.
Then, the other threat concerns the generalization of the results, as participants may not approximate well security administrators.
However, participants were selected on a voluntary basis, so they were motivated, they performed the task at their place with the most familiar environment, and they had unlimited time; therefore, in ideal conditions. 
Moreover, the ex post analysis of the correlation with the expertise didn't show inconsistencies.

More information about the experiments, the input policy and network, the observed data, the statistical assessment, and the threats to validity are available in the supplemental material of this paper.



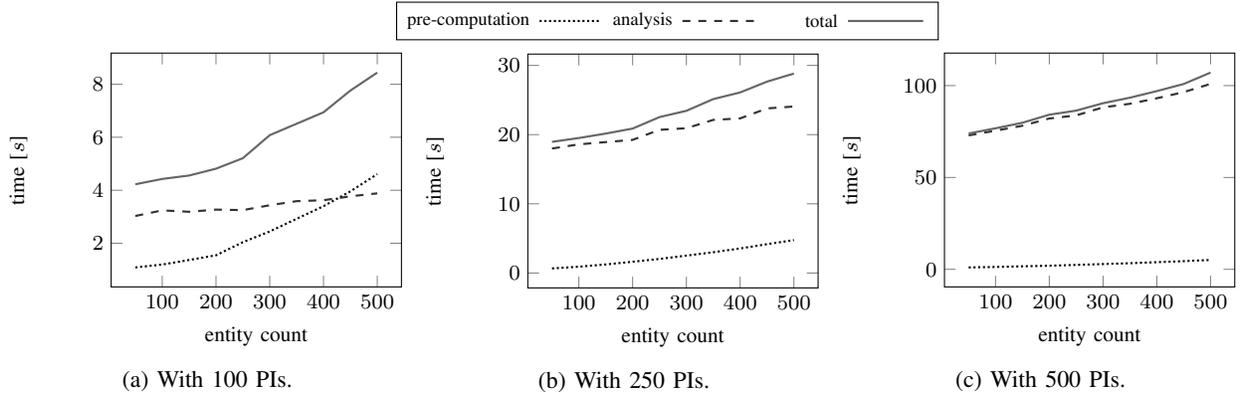
\begin{figure*}[t]
	\centering
	\begin{subfigure}{\textwidth}
		\centering
		\scalebox{1}{
			\begin{tikzpicture}
			\draw  (28.3,0.25) rectangle (35.5,-0.25);
			\draw [black, thick, densely dotted] (30.2,0)node[text=black,anchor=east,font=\scriptsize] {pre-computation} -- (31,0);
			\draw [black!80, thick, dashed] (32.1,0) node[text=black,anchor=east,font=\scriptsize] {analysis} -- (33.1,0);
			\draw [black!60, thick] (34.3,0) node[text=black,anchor=east,font=\scriptsize] {total} -- (35.3,0);
			\draw (27.3,-0.40) node[anchor=south, text=white] {.};
			\end{tikzpicture}
		}
	\end{subfigure}
	\footnotesize
	\begin{subfigure}{0.3\textwidth}
		\begin{tikzpicture}
		\begin{axis}[width = \textwidth, xlabel = {entity count}, ylabel = {time [$s$]}, xtick = {100, 200, 300, 400, 500}, mark size = 1.25, y filter/.code={\pgfmathparse{#1 / 1000}\pgfmathresult},
		legend style={
			legend pos=north west,
			font=\tiny, 
			legend columns=1
		},
		xlabel style={yshift=0.1cm},]

		\addplot[black, thick, densely dotted] table[x = Ent, y = tp, col sep = semicolon] {pis100.tex};
		\addplot[black!80, thick, dashed] table[x = Ent, y = tc, col sep = semicolon] {pis100.tex};
		\addplot[black!60, thick] table[x = Ent, y = tp+tc, col sep = semicolon] {pis100.tex};
		
		\end{axis}
		\end{tikzpicture}
		\caption{With 100 PIs.}
	\end{subfigure}
	\begin{subfigure}{0.3\textwidth}
		\begin{tikzpicture}
		\begin{axis}[width = \textwidth, xlabel = {entity count}, ylabel = {time [$s$]}, xtick = {100, 200, 300, 400, 500}, mark size = 1.25, y filter/.code={\pgfmathparse{#1 / 1000}\pgfmathresult}, xlabel style={yshift=0.1cm}, 
		legend style={
			legend pos=north west,
			font=\tiny, 
			legend columns=1
		},]
		\addplot[black, thick, densely dotted] table[x = Ent, y = tp, col sep = semicolon] {pis250.tex};
		\addplot[black!80, thick, dashed] table[x = Ent, y = tc, col sep = semicolon] {pis250.tex};
		\addplot[black!60, thick] table[x = Ent, y = tp+tc, col sep = semicolon] {pis250.tex};
		\end{axis}
		\end{tikzpicture}
		\caption{With 250 PIs.}
	\end{subfigure}
	\begin{subfigure}{0.3\textwidth}
		\begin{tikzpicture}
		\begin{axis}[width = \textwidth, xlabel = {entity count}, ylabel = {time [$s$]}, xtick = {100, 200, 300, 400, 500}, mark size = 1.25, y filter/.code={\pgfmathparse{#1 / 1000}\pgfmathresult}, xlabel style={yshift=0.1cm}]
		\addplot[black, thick, densely dotted] table[x = Ent, y = tp, col sep = semicolon] {pis500.tex};
		\addplot[black!80, thick, dashed] table[x = Ent, y = tc, col sep = semicolon] {pis500.tex};
		\addplot[black!60, thick] table[x = Ent, y = tp+tc, col sep = semicolon] {pis500.tex};
		\end{axis}
		\end{tikzpicture}
		\caption{With 500 PIs.}
	\end{subfigure}
	\caption{{Performance tests: time to perform the anomaly analysis of a fixed number of \PIs depending on the number of entities.}}
	\label{fig:PITests}
\end{figure*}
\begin{figure*}[t]
	\centering
	\begin{subfigure}{\textwidth}
		\centering
		\scalebox{1}{
			\begin{tikzpicture}
			\draw  (28.3,0.25) rectangle (35.5,-0.25);
			\draw [black, thick, densely dotted] (30.2,0)node[text=black,anchor=east,font=\scriptsize] {pre-computation} -- (31,0);
			\draw [black!80, thick, dashed] (32.1,0) node[text=black,anchor=east,font=\scriptsize] {analysis} -- (33.1,0);
			\draw [black!60, thick] (34.3,0) node[text=black,anchor=east,font=\scriptsize] {total} -- (35.3,0);
			\draw (27.3,-0.40) node[anchor=south, text=white] {.};
			
			\end{tikzpicture}
		}
	\end{subfigure}
	\footnotesize
	\begin{subfigure}{0.3\textwidth}
		\begin{tikzpicture}
		\begin{axis}[width = \textwidth, xlabel = {PI count}, ylabel = {time [$s$]}, xtick = {100, 200, 300, 400, 500}, mark size = 1.25, y filter/.code={\pgfmathparse{#1 / 1000}\pgfmathresult}, xlabel style={yshift=0.1cm}]
		\addplot[black, thick, densely dotted] table[x = PI, y = tp, col sep = semicolon] {entities100.tex};
		\addplot[black!80, thick, dashed] table[x = PI, y = tc, col sep = semicolon] {entities100.tex};
		\addplot[black!60, thick] table[x = PI, y = tp+tc, col sep = semicolon] {entities100.tex};
		\end{axis}
		\end{tikzpicture}
		\caption{With 100 entities.}
	\end{subfigure}
	\begin{subfigure}{0.3\textwidth}
		\begin{tikzpicture}
		\begin{axis}[width = \textwidth, xlabel = {PI count}, ylabel = {time [$s$]}, xtick = {100, 200, 300, 400, 500}, mark size = 1.25, y filter/.code={\pgfmathparse{#1 / 1000}\pgfmathresult}, xlabel style={yshift=0.1cm}]
		\addplot[black, thick, densely dotted] table[x = PI, y = tp, col sep = semicolon] {entities250.tex};
		\addplot[black!80, thick, dashed] table[x = PI, y = tc, col sep = semicolon] {entities250.tex};
		\addplot[black!60, thick] table[x = PI, y = tp+tc, col sep = semicolon] {entities250.tex};
		\end{axis}
		\end{tikzpicture}
		\caption{With 250 entities.}
	\end{subfigure}
	\begin{subfigure}{0.3\textwidth}
		\begin{tikzpicture}
		\begin{axis}[width = \textwidth, xlabel = {PI count}, ylabel = {time [$s$]}, xtick = {100, 200, 300, 400, 500}, mark size = 1.25, y filter/.code={\pgfmathparse{#1 / 1000}\pgfmathresult}, xlabel style={yshift=0.1cm}]
		\addplot[black, thick, densely dotted] table[x = PI, y = tp, col sep = semicolon] {entities500.tex};
		\addplot[black!80, thick, dashed] table[x = PI, y = tc, col sep = semicolon] {entities500.tex};
		\addplot[black!60, thick] table[x = PI, y = tp+tc, col sep = semicolon] {entities500.tex};
		\end{axis}
		\end{tikzpicture}
		\caption{With 500 entities.}
	\end{subfigure}
	\caption{{Performance tests: time to perform the anomaly analysis on networks of a fixed size depending on the number of \PIs.}}
	\label{fig:EntityTests}
\end{figure*}
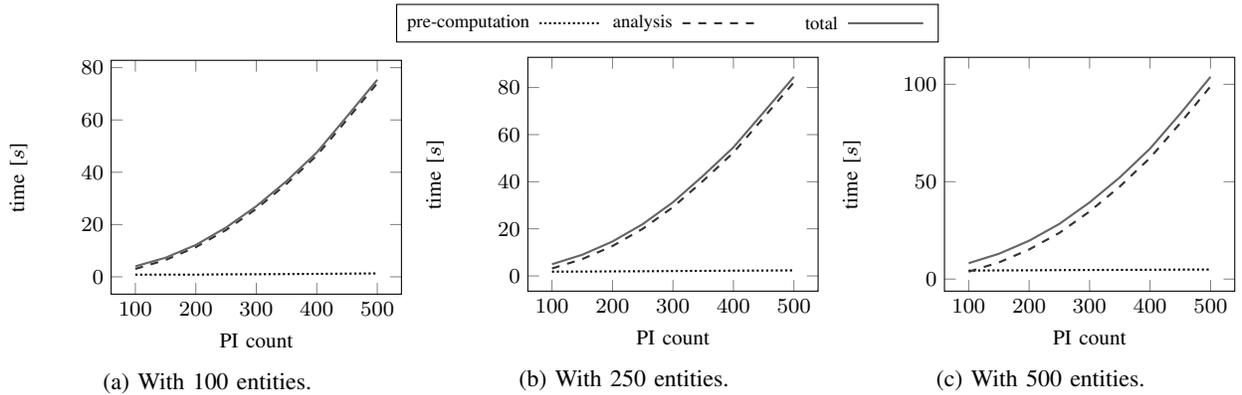

\subsection{Complexity analysis}
\label{sec:Complexity}

We will now derive some complexity formulas that prove the theoretical performance of our model. We will start with a simple observation. Our approach can be split in two consecutive phases. The first one is a \emph{pre-computation phase}, where the tree representation of the network and its paths are obtained. The second one is an \emph{analysis phase} that consists of the real anomaly detection phase.

Let's suppose that we have a network consisting of $\mathcal{E}$ entities (IPs, ports, addresses, \dots), $\mathcal{I}$ \policyImplementations and $\mathcal{C}$ connections between the network entities created by the \PIs (obviously $\mathcal{C} \ge \mathcal{I}$).

We will start by taking a look at the pre-computation phase. To create the tree representation of the network nodes, we need to check every single entity, so that this process has an exact complexity of $\mathcal{E}$. Finding all the simple paths\footnote{A simple path is a path with no duplicate vertexes.} in an acyclic graph is a \textsc{NExpTime} problem with a maximum complexity of $O(e^\mathcal{E})$. Note, however, that the real networks are scarcely connected and that multiple paths between two different nodes are quite rare, making these calculations feasible also in large IT infrastructures. In addition, an administrator can choose to limit the number of paths to check to some fixed value $\mathcal{P}$, typically $\mathcal{P} \lll e^\mathcal{E}$. Hence, the total complexity of the pre-computation phase is $\mathcal{E} + \mathcal{P}$.

Regarding the analysis phase, we have to take into account the different characteristics of the anomaly detection formulas. In particular, we have that:
\begin{itemize}
	\item the \internalLoop, \outOfPlace, \nonEnforceability, \inadequacy, \filteredChannel, \LTwo and \asymmetricChannel anomalies algorithms work on a single \PI at a time, so that they have a complexity of $\mathcal{I}$;
	\item the \shadowing, \redundancy, \exception and \inclusion anomalies algorithms need an ordered pair of \PIs, thus have a complexity of $\mathcal{I} (\mathcal{I} - 1)$. Also the \superfluous anomaly has a quadratic complexity since it needs to test every \PI against all the other ones;
	\item the \correlation, \affinity and \skewedChannel anomalies algorithms work on unordered pairs of \PIs, giving a complexity of $\mathcal{I} (\mathcal{I} - 1) / 2$;
	\item the \monitorability and \alternativePath anomalies algorithms work on every path, hence their complexity is $\mathcal{P}$;
	\item the \cyclicPath anomaly algorithm can be efficiently performed using a proper cycle detection algorithm such as~\cite{Tarjan72}, that has a complexity of $O(\mathcal{E} + \mathcal{C})$. Note that its complexity is not necessarily $\mathcal{P}$ since a graph with few loops may have an infinite number of paths.
\end{itemize}
Summarizing, the total complexity of the analysis phase is:
\begin{equation*}
\mathcal{I} + \mathcal{I} (\mathcal{I} - 1) + \mathcal{P} + O(\mathcal{E} + \mathcal{C}) \approx \mathcal{I}^2 + \mathcal{P} + O(\mathcal{E} + \mathcal{C})
\end{equation*}

\subsection{Performance analysis}
\label{sec:Performances}

We implemented our anomaly detection model and tested it in several scenarios using a number of synthetically generated networks in order to assess its running time.
Our tool was developed using Java 1.8 and the natural graph-based representation of ontologies offered by OWL API 3.4.10 and the reasoner Pellet 2.3.1. We performed all our tests on an Intel i7 @ \SI{2.4}{\giga\hertz} with \SI{16}{\giga\byte} RAM under Windows~7.

Each test was performed on several ad-hoc scenarios consisting of an automatically generated network with a parametric structure where we can specify:
\begin{enumerate*}
	\item the number of network entities;
	\item the number of \policyImplementations;
	\item the percentage of conflicting \PIs.
\end{enumerate*}
We choose to fix the number of conflicting \PIs to about $50\%$ since, from our empirical analysis, on average, an administrator writes only about half of the \policyImplementations without any kind of conflict (Table~\ref{Tab2a}).
We performed two kinds of tests. 
In the first one we fixed the number of network entities and increased the number of \PIs, while in the second one we did the reverse (fixed the number of \PIs and changed the entities count).
Fig.~\ref{fig:PITests} shows the test results when fixing the number of \PIs respectively to $100$, $250$ and $500$, while Fig.~\ref{fig:EntityTests} shows the graphs when the network entities count is $100$, $250$ and $500$.
We kept track of three times: the pre-computation (the dotted lines), the analysis phase (the dashed lines) and the total times (the solid lines).

Our tool proved to be very scalable, achieving a total time of less than two minutes in the worst scenario (500 \PIs and 500 network entities).
We also noted that the results are aligned with the complexity analysis in Section~\ref{sec:Complexity}. 
For instance, the computation times grow when the number of network entities (Fig.~\ref{fig:PITests}) increase, while the pre-computation phase time is independent of the number of entities (Fig.~\ref{fig:EntityTests}).

\section{Related works}
\label{sec:RelatedWorks}

Anomaly analysis, detection and resolution in policy-based systems and security controls are hot topics.
We briefly discuss in the next paragraphs the most notable works in this area.

\subsection{Communication protection policies}

The current literature contains several works about anomaly detection in \cpp, however, the research in this area is solely focused on IPsec, and overlooks the effects of multiple overlapping protection techniques.

Zao~\cite{Zao00} introduced an approach based on the combination of conditions that belong to different IPsec fields.
The same idea was used by Fu\etal \cite{Fu01} to describe a number of conflicts between IPsec tunnels, discovered through a simulation process that reports security requirements violations.
In their analysis, the policy anomalies are identified by checking the IPsec configurations against the desired policies written in a natural language.
In practice, an anomaly occurs when the policy implementations do not satisfy the desired policies. 
In addition, Fu\etal proposed a resolution process that finds alternative configurations to satisfy the desired policy.

Hamed\etal's analyzed the effects of IPsec rules on the protection of networks~\cite{AlIPsec}, by proposing a number of ad-hoc algorithms and formulas to detect these problems. They formalized the classification scheme of~\cite{Fu01} and proposed a model based on OBDD (Ordered Binary Decision Diagrams) that not only incorporates the encryption capabilities of IPsec, but also its packet filter capabilities. He also identified two new IPsec problems (channel-overlapping and multi-transform anomalies). The first one occurs when multiple IPsec sessions are established and the second session redirect the traffic of the first one (similar to the case depicted in Fig.~\ref{fig:TrellisDiagram}). On the other hand, the multi-transform anomalies occur when a data protection is applied to an already encapsulated IPsec traffic and the second protection is weaker than the first one. The same authors also described a classification system for conflicts between filtering and communication protection policies~\cite{AlTax}.
Niksefat \etal \cite{Niksefat2010_IPSec} presented an two improvements over the Hamed\etal's solution \cite{AlIPsec}, a faster detection algorithm and the possibility to resolve the detected anomalies.
Finally, Li \etal classified the IPsec rules in two classes: access control lists (ACL) and encryption lists (EL) \cite{Li06}.
\subsection{Filtering policies}

Configuration/policy anomaly detection in networks is not only restricted to communication protection technologies. In literature there exist a rich collection of papers about filtering policy analysis. These works belong to a different domain w.r.t. the approach presented in this paper, however they are interesting background on anomaly analysis. Note that in our approach we perform a full-spectrum CPP analysis that takes also into account the effects of filtering through the filtered channel anomaly, while the works presented in the following paragraphs focus solely on the conflicts between filtering rules.

One of the most influential works in this area is by Al-Shaer\etal, which addresses the analysis of filtering configurations via FOL formulas \cite{Al-Shaer2005}. The authors analyzed both the local anomalies arising in a single firewall and the global ones taking into account several distributed filtering devices.

Liu\etal focused on detecting and removing redundant filtering rules with data-structure named FDD (Firewall Decision Diagram) \cite{Liu2005_RedundancyFirewalls}. The authors distinguish upward redundant rules, which are rules that are never matched, and downward redundant rules, which are rules that are matched but enforce the same action as some lower priority rules.

Basile\etal described a geometric representation of filtering rules (up to the application level) where detection and resolution is based on the intersection of hyper-rectangles \cite{Basile12,basile2015ton,basile2016}.
The authors extended the work performed by Al-Shaer by introducing the anomalies between more than two rules and by showing how to transform a policy representation in another form that preserves its semantic.
Similarly, Hu\etal suggested to split the classic five-tuple decision space into disjoint hyper-rectangles, induced by the rules, where conflicts are resolved with a combination of automatic strategies \cite{Hu12}.

Hu\etal proposed a thoroughly different approach, an ontology-based management framework that detects anomalies in filtering rule sets \cite{Hu11}.
Analogously, Bandara\etal proposed the use of logic reasoning, obtaining excellent performance \cite{Bandara09}.
Alfaro\etal presented a collection of algorithms to remove a series of anomalies between packet filter configurations and NIDS in distributed systems \cite{Alfaro08} that have been implemented in the MIRAGE tool \cite{Alfaro10}.


\section{Conclusions and future work}
\label{sec:Conclusions}

In this paper we proposed a list of nineteen anomalies (classified in two taxonomies) that can arise during the implementation of communication protection policies, and devised a formal model based on FOL formulas that is able to detect them.
Our approach can be used to find incompatibilities, redundancies and severe errors among policy implementations that use security technologies working at different OSI layers and with different security properties.

We implemented our model in Java by making an extensive use of ontological techniques, and verified in several network scenarios that it is scalable and performs well.
Most of the anomalies can be visualized with a graph-based representation that should facilitate their identification.



For the future, we plan to extend the expressivity of our model by adding support for new types of network devices, such as intrusion detection systems (IDS). Furthermore, we are planning to perform other empirical assessments to evaluate if our tool can help administrators to reduce the number of anomalies in real-world scenarios.





%
%
%
\vspace*{-1.3cm}
\begin{IEEEbiography}[{\includegraphics[width=1in,height=1.25in,clip,keepaspectratio]{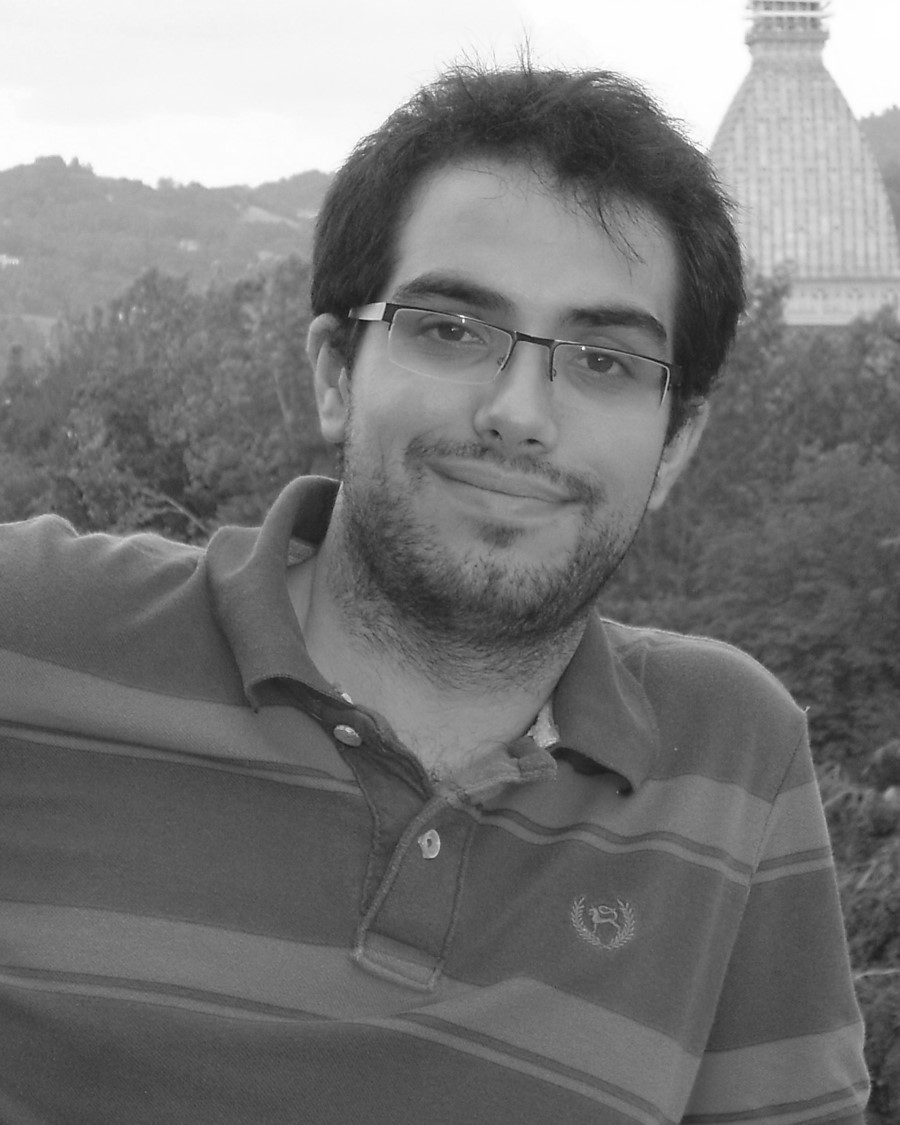}}]{Fulvio Valenza}
	received the M.Sc. degree (summa cum laude)  in computer engineering  in 2013 from the Politecnico di Torino. In 2016, he completed his PhD in Computer Engineering at the Politecnico di Torino. His research activity focused on network security policies. Currently he is a Researcher at the CNR-IEII Torino, Italy, where he works on orchestration and management of network security functions in the context of SDN/NFV-based networks and industrial systems.
	
\end{IEEEbiography}

\vspace*{-1.2cm}

\begin{IEEEbiography}[{\includegraphics[width=1in,height=1.25in,clip,keepaspectratio]{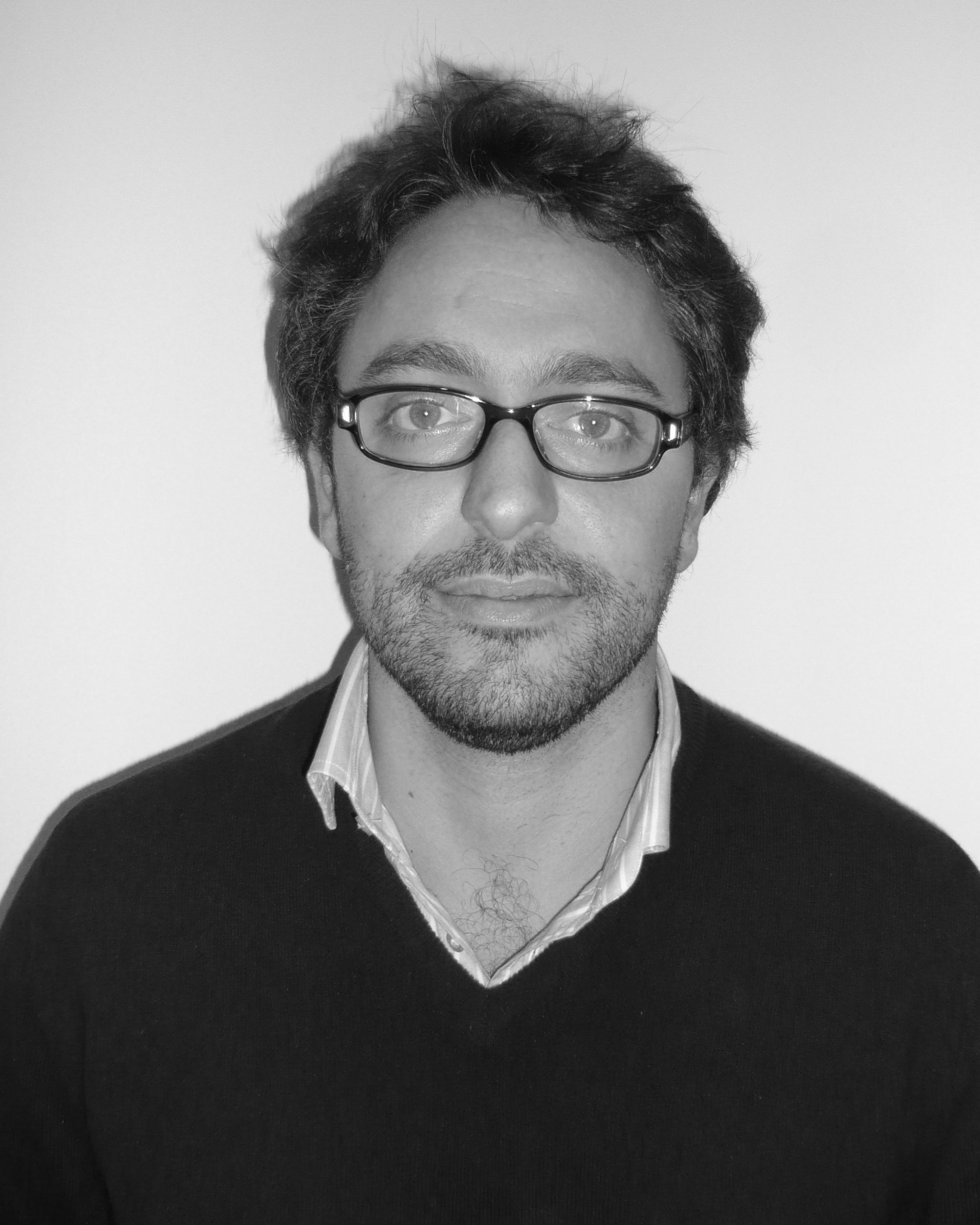}}]{Cataldo Basile}
	received a M.Sc. (summa cum laude) in 2001 and a Ph.D. in Computer Engineering in 2005 from the Politecnico di Torino, where is currently a research assistant. His research is concerned with policy-based management of security in networked environments, policy refinement, general models for detection, resolution and reconciliation of specification conflicts, and software security.
\end{IEEEbiography}
%
%
\vspace*{-1.5cm}
\begin{IEEEbiography}[{\includegraphics[width=1in,height=1.25in,clip,keepaspectratio]{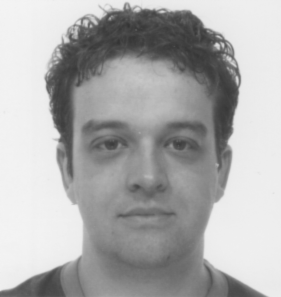}}]{Daniele Canavese}
	received the M.Sc. degree in computer engineering in 2010 from the Politecnico di Torino, where he is currently a research assistant and a Ph.D. student. His research interests are concerned with policy-based management systems, models for network analysis, security management via inferential systems and public key cryptography.
\end{IEEEbiography}
\vspace*{-1.5cm}
\begin{IEEEbiography}[{\includegraphics[width=1in,height=1.25in,clip,keepaspectratio]{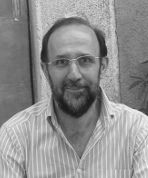}}]{Antonio Lioy}
	is full professor at the Politecnico di Torino, where he leads the TORSEC research group active in information system security. His research interests include network security, policy-based system protection, and electronic identity. Lioy received a M.Sc. in Electronic Engineering (summa cum laude) and a Ph.D. in Computer Engineering, both from the Politecnico di Torino.
\end{IEEEbiography}

%
%

\newpage
\appendix
In this supplemental document, we provide additional material regarding our model, its implementation, and validation. In particular:
\begin{itemize}
	\item in Section~\ref{sec:empirical} we provide the exercise undertaken by the participants of our empirical assessment, we report the data analysis to prove the statistical significance of the inferred statements, and sketch the threats to validity;
	\item in Section~\ref{sec:mapping} we present the configurations of some selected communication protection security controls and their mapping into our model as a set of \policyImplementations to prove the expressiveness of our approach and the easiness of the translation;
	\item in Section~\ref{sec:net_gen} we describe the algorithms to generate the synthetic networks used to executed the performance and scalability tests;
	\item in Section~\ref{sec:notations} we summarize in a table all the symbols and notations used in the paper.
\end{itemize}

\subsection{Empirical assessment}\label{sec:empirical}
This section contains additional material regarding the empirical assessment of our model.
We first provide a complete description of the exercise given to the test population, then we present the analysis of variances performed to validate the significance of our results for the second research question (the higher the expertise of the administrators the lower the number of introduced anomalies).

\subsubsection{Questionnaire}
We asked the subjects to translate a set of five high level IT communication protection policies into a set of configuration parameters included in our notion of policy implementations (PIs).
We recall here that in our model, a \PI $i$ is a tuple\footnote{All these parameters are explained in detail in Section~V of the paper.}:
\begin{equation*}
i = (s, d, t, C, S, G)
\end{equation*}
where:

\begin{itemize}
	\item $s$ and $d$ respectively represent the channel source and destination;
	\item $t$ is the adopted security technology;
	\item $C$ is an ordered set of coefficients that indicate the required security levels;
	\item {$S$ is a tuple of selectors used to identify the traffic that need to be protected;}
	\item $G$ is the list of gateways involved in the communication.
\end{itemize}

We posed no limits on the number of PIs and on the time required to complete the experiment. 

To train the participants, we provided a brief description of the PI model, a small network example and a few examples of \policyImplementations on the example network.
For the sake of brevity we omit here the training examples and only report the exercise given to the subjects.

The material provided to the subjects before starting the exercise included:
\begin{enumerate*}
	\item the topology graph of the network where to enforce the communication protection policies;
	\item the explanation of all the services available at each graph node;
	\item a set of high level policies;
	\item a set of constraints in the policy enforcement.
\end{enumerate*}

All this material was available both online, on the web form where we asked their contribution, and as a set of downloadable pdf files to be read off-line.
Subjects were asked to fill in their PIs on a web form that constrained them to use valid values.

\begin{figure*}
	\centering
	\scalebox{1}{
		\noindent\fbox{%
			\parbox{\columnwidth}{%
				\textbf{User groups}: network nodes where they can connect
				\begin{itemize}
					\item \textit{DB admins}: $C_{C1}$, $C_{C2}$
					\item \textit{accounting service admins}: $C_{C2}$, $C_{C3}$
					\item \textit{admins}: $C_{C1}$,$C_{C2}$, $C_{C3}$
					\item \textit{Turin employees}: $C_{A1}$, $C_{A2}$, $C_{A3}$
					\item \textit{Palermo employees}: $C_{D1}$, $C_{D2}$, $C_{D3}$
					\item \textit{employees}: $C_{A1}$, $C_{A2}$, $C_{A2}$, $C_{D1}$, $C_{D2}$, $C_{D3}$
					
				\end{itemize}
				\noindent\textbf{Services}: network nodes where they are accessible
				\begin{itemize}
					\item \textit{Database}: $S_{C1}$, $S_{B1}$
					\item \textit{Accounting service}: $S_{C1}$, $S_{B1}$
					\item \textit{Accounting admin front end}: $S_{C1}$, $S_{B2}$
					\item \textit{Mail service}: $S_{E1}$, $S_{E2}$
				\end{itemize}
				\noindent\textbf{Constraints and Requirements}
				\begin{itemize}
					\item $G_{E1}$ filters all the encrypted traffic to/from $S_{E1}$ and $S_{E2}$
					\item $S_{B1}$ only supports IPSec and TLS.
					\item $S_{B2}$ only supports TLS and SSH.
					\item $S_{C1}$ only supports IPSec and SSH.
					\item if not specified, every node supports all the communication protection technologies (IPSec, TLS, SSH and WS-Security).
					\item all the communications between entities in two different corporate subnets must be protected against sniffing from non corporate people.
					\item enforce confidentiality when a connection is crossing $G_1$, $G_2$ or $G_3$.
					\item confidentiality is optional if a connection is within a single subnet.
				\end{itemize}
			}
			\parbox{\columnwidth}{%
				\scalebox{1}{
					{\includegraphics[width=1\linewidth]{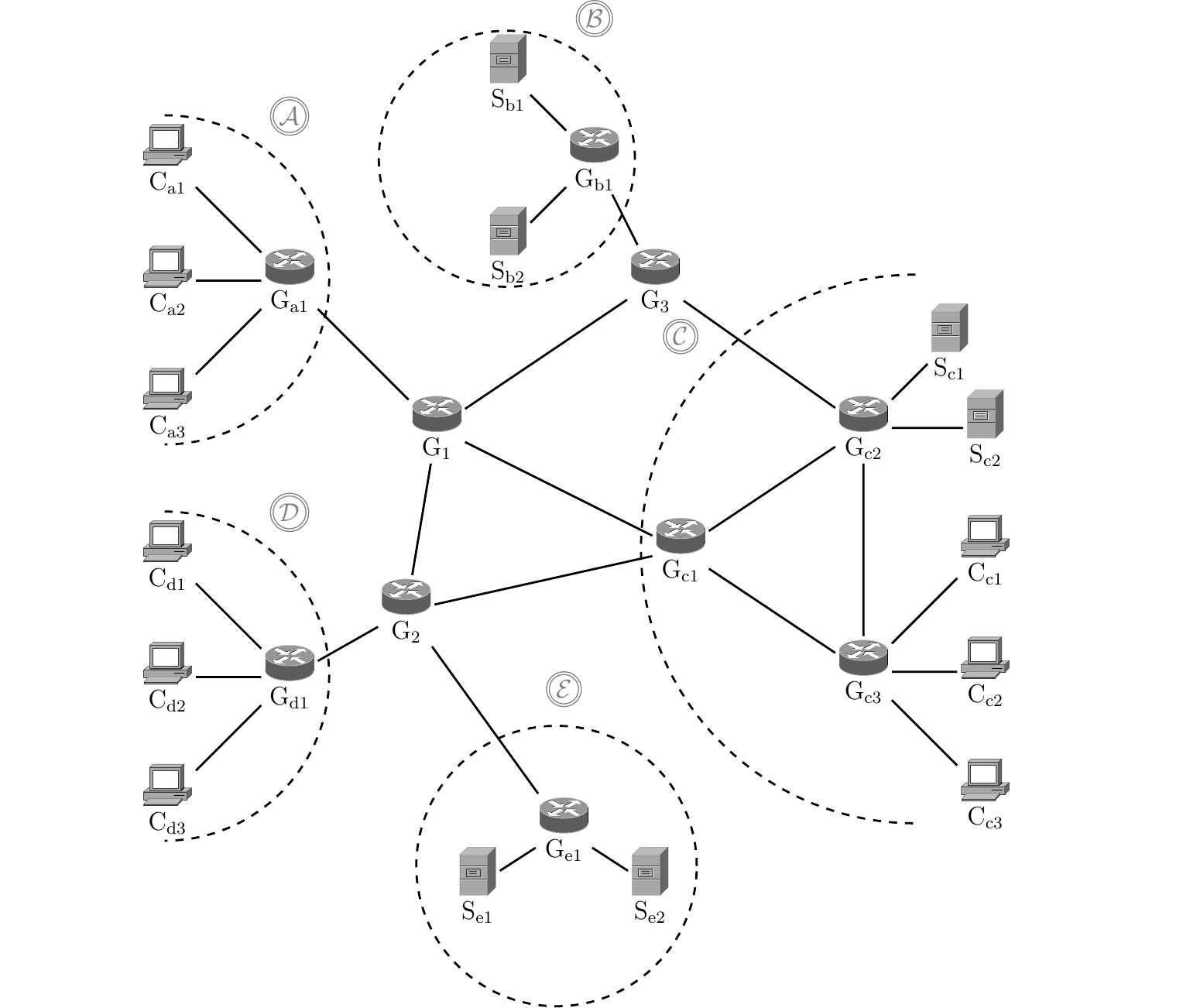}}
				}

				\noindent\textbf{High level IT Policies}
				\begin{enumerate}
					\item the accounting services securely reach the Databases
					\item the DB admins securely reach the Databases
					\item the accounting service  admins securely reach the accounting admin front end
					\item all the admins securely reach the accounting services
					\item all the employees reach the mail service
				\end{enumerate}
				
				Note: tn this context, securely reaches means that we care about the security of the communication (that is choose at your discretion the technology and the security properties to enforce). On the other hand, reaches means we do not care about security (so a secure or insecure connection is fine).

			}%
		}
	}
	\caption{Material provided to the subjects.}
	\label{fig:network}
\end{figure*}

The network topology graph, the additional information about network entities and users are shown in Fig.~\ref{fig:network}.

\subsubsection{Analysis of variances}
To prove the research question RQ2 (``does the number of anomalies decrease when the administrator expertise grows?''), we divided the subjects in three sets based on their expertise. We prove here, via a one-way analysis of variances (one-way ANOVA), that the expertise level (high, medium and low) is a significant factor in introducing anomalies.

We assume that a hypothesis is valid if the significance level is less than $0.05$ (the most commonly used threshold).

ANOVA tests the non-specific null hypothesis ($H_0$) that all three population means are equal, that is:

\begin{equation*}
H_0: \mu_{higt}=\mu_{medium}=\mu_{low}
\end{equation*}
On the other hand, the alternative hypotheses is obviously:
\begin{equation*}
H_a: H_0 \text{ is false}
\end{equation*}
We performed the ANOVA analysis on two different data sets:
\begin{itemize}
	\item the number of anomalies introduced by each administrator (test 1);
	\item the number of anomaly types introduced by each administrator (test 2).
\end{itemize}

The computed P-values of the ANOVA are 0.034 (test1) and 0.006 (test2). In both cases, being the P-values less than the significance level, the null hypothesis can be rejected, thus the population means are not all equal when introducing anomalies. In addition both the F statistic numbers are greater than their minimum F critical values, further contributing to accepting the alternative hypothesis \cite{anova}.

In the following sections we report more details about our analysis.

\subsubsection{Test 1 -- anomalies for each administrator}

In Table~\ref{tab:data11} we present the number of anomalies introduced by each administrator depending on their expertise.
Administrators have been divided in $l_1$-$l_{10}$, $m_1$-$m_{10}$ and $h_1$-$h_{10}$.
Each cell value represents the number of anomalies introduced by an administrator.

Table~\ref{tab:stats1} shows the statistical results computed on these data, while Table~\ref{tab:anova1} details the ANOVA computation values. The table columns have the following meanings:

\begin{itemize}
	\item SS stands for `Sum of Squares' and it is the sum of squared deviations;
	\item df is the acronym for `Degree of Freedom';
	\item MS is the `Mean Square`, a sort of standard deviation;
	\item F is the `F statistic', a value used to accept the alternative hypothesis if it is large enough;
	\item P-value is the `Probability' of having an F statistic	larger enough such that	the	null hypothesis	is true;
	\item F crit is the `F critical value', used to accept the alternative hypothesis if F statistic is greater than this threshold.
\end{itemize}

\subsubsection{Test2 -- anomaly types for each administrator}

Table~\ref{tab:data21} presents the number of anomaly types for which the administrators introduced at least one anomaly, split according to their expertise level.
Tables~\ref{tab:stats2}~and~\ref{tab:anova2} show the statistical results computed on these data.

\begin{table}[h]
	
	\resizebox{\columnwidth}{!}
	{%
	\centering
	\begin{tabular}{ccccccccccc}
		 \multicolumn{11}{r}{high expertise administrators}\\ 
		\toprule
        admin& $h_1$&$h_2$&$h_3$&$h_4$&$h_5$&$h_6$&$h_7$&$h_8$&$h_9$&$h_{10}$\\
        \midrule
        anomaly \# & 1&2&2&4&10&7&0&2&5&7\\
		\bottomrule
	\end{tabular}
}
\medskip
\medskip

	\resizebox{\columnwidth}{!}{%
\begin{tabular}{ccccccccccc}
			 \multicolumn{11}{r}{medium expertise administrators}\\ 
	\toprule
	admin& $m_1$&$m_2$&$m_3$&$m_4$&$m_5$&$m_6$&$m_7$&$m_8$&$m_9$&$m_{10}$\\
	\midrule
	anomaly \# & 6&7&6&2&3&0&7&6&6&22\\
	\bottomrule
\end{tabular}
}

\medskip
\medskip

	\resizebox{\columnwidth}{!}{%
	\begin{tabular}{ccccccccccc}
	\multicolumn{11}{r}{low expertise administrators}\\ 
	\toprule
	admin& $l_1$&$l_2$&$l_3$&$l_4$&$l_5$&$l_6$&$l_7$&$l_8$&$l_9$&$l_{10}$\\
	\midrule
	anomaly \# & 4&20&24&4&20&19&13&4&9&12\\
	\bottomrule
\end{tabular}
}
	\caption{Anomalies count per administrators (test 1).}
	\label{tab:data11}
\end{table}

\begin{table}[h!]
	\centering
	\resizebox{\columnwidth}{!}
	{%
		\begin{tabular}{lcccc}
			\toprule
			expertise & admins  & anomalies sum & average anomalies & variance\\
			\midrule
			high & 10    & 40  & 4       & 10.222 \\
			medium & 10    & 65  & 6.5     & 35.167 \\
			low & 10    & 129 & 12.9    & 57.211 \\
			\bottomrule
		\end{tabular}
	}
	\caption{Statistics for the test 1.}
	\label{tab:stats1}
\end{table}
\begin{table}[h!]
	\centering
	\begin{tabular}{lllllll}
		\toprule
		source of variation & SS     & df & MS    & F        & P-value & F crit\\
		\midrule
		between groups      & 421.4  & 2  & 210.7 & 6.161 & \colorbox{Black!20}{0.006} & 3.354 \\
		within groups       & 923.4  & 27 & 34.2  &          &         &          \\
		\bottomrule
	\end{tabular}
	\caption{ANOVA for the test 1.}
	\label{tab:anova1}
\end{table}

%
%
%

\begin{table}[h]
	\centering
		\resizebox{\columnwidth}{!}{%
	\begin{tabular}{ccccccccccc}
				 \multicolumn{11}{r}{high expertise administrators}\\ 
		\toprule
        admin& $h_1$&$h_2$&$h_3$&$h_4$&$h_5$&$h_6$&$h_7$&$h_8$&$h_9$&$h_{10}$\\
        \midrule
        anomaly \# & 1&1&1&2&2&2&0&2&3&2\\
		\bottomrule
	\end{tabular}
}
\medskip
\medskip

	\resizebox{\columnwidth}{!}{%
	\begin{tabular}{ccccccccccc}
						 \multicolumn{11}{r}{medium expertise administrators}\\ 
	\toprule
	admin& $m_1$&$m_2$&$m_3$&$m_4$&$m_5$&$m_6$&$m_7$&$m_8$&$m_9$&$m_{10}$\\
	\midrule
	anomaly \# & 1&3&1&2&2&0&3&2&3&4\\
	\bottomrule
\end{tabular}
}
\medskip
\medskip

	\resizebox{\columnwidth}{!}{%
	\begin{tabular}{ccccccccccc}
						 \multicolumn{11}{r}{low expertise administrators}\\ 
	\toprule
	admin& $l_1$&$l_2$&$l_3$&$l_4$&$l_5$&$l_6$&$l_7$&$l_8$&$l_9$&$l_{10}$\\
	\midrule
	anomaly \# & 3&4&4&3&1&3&3&1&3&4 \quad\\
	\bottomrule
\end{tabular}
}
	\caption{Anomaly types count per administrators (test 2).}
	\label{tab:data21}
\end{table}

\begin{table}[h!]
	\centering
	\resizebox{\columnwidth}{!}
	{%
		\begin{tabular}{lcccc}
			\toprule
			expertise & admins &  anomaly types sum & average type & variance\\
			\midrule
			high & 10    & 16  & 1.6       & 0.711 \\
			medium & 10    & 21  & 2.1     & 1.433 \\
			low & 10    & 29 & 2.9    	   & 1.211 \\
			\bottomrule
		\end{tabular}
	}
	\caption{Statistics for the test 2.}
	\label{tab:stats2}
\end{table}

\begin{table}[h!]
	\centering
	\begin{tabular}{lllllll}
		\toprule
		source of variation & SS     & df & MS    & F        & P-value & F crit\\
		\midrule
		between groups      &8.6  & 2  & 4.3& 3.84 & \colorbox{Black!20}{0.034} & 3.354 \\
		within groups       & 30.2  & 27 & 1.119  &          &         &          \\
		\bottomrule
	\end{tabular}
	\caption{ANOVA for the test 2.}
	\label{tab:anova2}
\end{table}

%
%
%

\subsubsection{Threats to validity}

We have designed a simple experiment focused on a single purpose, \ie verifying the possibility to introduce anomalies when writing \cpp, with a clear causal inference.

Nevertheless, we have analysed the validity of our experiment against the threat proposed by Wohlin \etal \cite{wohlin00}. 
In summary, threats (construct, internal, external, conclusion) are limited because all the used metrics were based on objective data (counting anomalies and their types), the test was anonymous and performed off-line, the experiment was not ambiguous (according to participants), and data were collected with questionnaires and designed according to standard methods and scales and analysed with state of the art statistical methods.
However, we identified two major threats.
First, for practical reasons (anonymity, expected number of participants, web-based data collection) we provided a single administration task, therefore, the experiment is subject to the mono-operational bias. 
This threat has been mitigated as the activities to perform were various, realistic, and complex enough to capture the heterogeneity of administrators' tasks.
Then, we have issues on the generalization of the results, as participants may not approximate well security administrators.
However, we designed the experiment to mitigates this risk.
Participants were selected on a voluntary basis, so they were motivated. They performed the task at their places, with the most familiar environment, and they had unlimited time, therefore, in ideal conditions. Moreover, they were classified based on their expertise.
Other generalization threats related to the network size and complexity, which was designed to be complex enough but to allow completing the task in one hour, and the policies were real security requirements.

\subsection{Mapping \policyImplementations to technology-specific configurations}\label{sec:mapping}
\lstset{frameround=fttt}

Translating data and channel protection configurations into/from the policy representations we use in our model is a straightforward task.
The \policyImplementations convey in a very compact way all the information that is needed to correctly configure the security controls.

To prove our claim, we provide the mapping of configurations of three different technologies:
\begin{itemize}
	\item IPsec, by mapping a strongSwan\footnote{See \url{https://www.strongswan.org/}.} configuration for an end-to-end, site-to-site and remote access channels;
	\item TLS, by mapping an OpenVPN\footnote{See \url{https://openvpn.net/index.php/open-source/documentation/}.} configuration to establish a secure tunnel;
	\item SSH, by mapping a SSH tunnel configuration.
\end{itemize}
Analogously, all the configurations of the secure communications used in our analysis model can be mapped into \policyImplementations.

\subsubsection{IPsec configurations (strongSwan)}
Listing~\ref{h2h} shows a strongSwan end-to-end connection from the host 192.168.0.100 to the IP address 192.168.0.200, which requires aes256-sha512-modp2048 as ESP parameters.
\begin{lstlisting}[captionpos=b,caption={End-to-end strongSwan configuration.},label=h2h, language=XML,frame=single,   basicstyle=\ttfamily\small,  xleftmargin=20pt,
xrightmargin=20pt,]  % Start your code-block

config setup
conn %default
ikelifetime=60m
keylife=20m
rekeymargin=3m
keyingtries=1
keyexchange=ikev1
conn host-host
left=192.168.0.100
leftcert=moonCert.pem
leftid=@moon.strongswan.org
leftfirewall=yes
right=192.168.0.200
rightid=@sun.strongswan.org
ike=aes256-sha512-modp2048
esp=aes256-sha512-modp2048
\end{lstlisting}

This configuration is summarized by the following \PI:
\begin{align*}
( 192.168.0.1, 192.168.0.2, \textrm{IPsec}, (5, 5, 5), *,\varnothing )
\end{align*}

\medskip
Listing~\ref{s2s} presents a site-to-site configuration between the subnets 10.1.0.0/16 and 10.2.0.0/16 operated by the gateways whose IP addresses are 192.168.0.1 and 192.168.0.2, respectively.
\begin{lstlisting}[captionpos=b,caption={Site-to-site strongSwan configuration.},label=s2s, language=XML,frame=single,   basicstyle=\ttfamily\small,  xleftmargin=20pt,
xrightmargin=20pt,]

conn %default
ikelifetime=60m
keylife=20m
rekeymargin=3m
keyingtries=1
keyexchange=ikev1
conn net-net
left=192.168.0.1
leftcert=moonCert.pem
leftid=@moon.strongswan.org
leftsubnet=10.1.0.0/16
leftfirewall=yes
right=192.168.0.2
rightid=@sun.strongswan.org
rightsubnet=10.2.0.0/16
ike=aes256-sha512-modp2048
esp=aes256-sha512-modp2048
\end{lstlisting}

This configuration can be represented with the following \PI:

\begin{align*}
& ( 192.168.0.1, 192.168.0.2, \textrm{IPsec}, (5, 5, 5),\\
&\;\; (10.1.0.0/16, *, 10.2.0.0/16, *, * ),\varnothing )
\end{align*}

Finally, Listing~\ref{ra} configures a remote access for the host 192.168.0.100 to allow access the subnet 10.2.0.0/16 through the gateway 192.168.0.1.

\begin{lstlisting}[captionpos=b,caption={Remote access strongSwan configuration.},label=ra, language=XML,frame=single,
basicstyle=\ttfamily\small,  xleftmargin=20pt,
xrightmargin=20pt,]  % Start your code-block

conn %default
ikelifetime=60m
keylife=20m
rekeymargin=3m
keyingtries=1
keyexchange=ikev1
conn home
left=192.168.0.100
leftsourceip=%config
leftcert=carolCert.pem
leftid=carol@strongswan.org
leftfirewall=yes
right=192.168.0.1
rightsubnet=10.1.0.0/16
rightid=@moon.strongswan.org
ike=aes256-sha512-modp2048
esp=aes256-sha512-modp2048
\end{lstlisting}
The \policyImplementation for this configuration is:
\begin{align*}
&( 192.168.0.100, 192.168.0.1, \textrm{IPsec}, (5, 5, 5),  \\
&\quad\quad\quad(*, *, 10.2.0.0/16, *, *),\varnothing )
\end{align*}

\medskip

\subsubsection{TLS (OpenVPN)}
Listing~\ref{tls} shows a configuration for OpenVPN of a tunnel from a generic client (\ie 192.168.1.100) to the server 192.168.1.1:1194, which requires the enforcement of a cipher-suite that uses Diffie-Hellman with AES-256-CBC and SHA512.

\begin{lstlisting}[captionpos=b,caption={Client side OpenVPN 2.0 configuration.},label=tls, language=XML,frame=single, basicstyle=\ttfamily\small,  xleftmargin=20pt,
xrightmargin=20pt,]  % Start your code-block

client
dev tun
proto udp
remote my-server 192.168.1.1:1194
nobind
ca ca.crt
cert client.crt
key client.key
remote-cert-tls server
tls-auth ta.key 1
cipher AES-256-CBC
auth SHA-512
dh dh1024.pem

\end{lstlisting}
This configuration translates into the following \PI:
\begin{align*}
(192.168.1.100 :*, 192.168.1.1\textrm{:}1194, \textrm{TLS}, (5, 5, 5),*,\varnothing )
\end{align*}
{Listing~\ref{tlsServer} shows the server relative OpenVPN configuration.}

\begin{lstlisting}[captionpos=b,caption={Server side OpenVPN 2.0 configuration.},label=tlsServer, language=XML,frame=single, basicstyle=\ttfamily\small,  xleftmargin=20pt,
xrightmargin=20pt,]  % Start your code-block

local 192.168.1.1
port 1194
dev tun
proto udp
keepalive 10 120
ca ca.crt
cert server.crt
key server.key
tls-auth ta.key 0
cipher AES-256-CBC
auth SHA-512
dh dh1024.pem
\end{lstlisting}

\subsubsection{SSH}
Listing~\ref{ssh} presents the configuration of an SSH tunnel for the user whose username is `client' and its IP address is 192.168.2.100. The user connects to the SSH server at 192.168.2.1:22022, which uses AES256-CBC and SHA512 to secure the data and enters the local network 10.0.0.0/16 with the 10.0.0.3 IP address on port 3306.

\begin{lstlisting}[captionpos=b,caption={Client side SSH configuration.},label=ssh, language=XML,frame=single, basicstyle=\ttfamily\small,  xleftmargin=15pt,
xrightmargin=10pt,]  % Start your code-block

Host tunnel
#SSH connection setting
HostName 192.168.2.1
User client
Port 22022
IdentityFile ~/.ssh/client.example.key
Ciphers aes256-cbc
MACs hmac-sha2-512 

#SSH tunnel setting
LocalForward 10.0.0.3:3306 127.0.0.0:3306
\end{lstlisting}

The \PI corresponding to this SSH configuration is:
\begin{align*}
&( 192.168.2.100:*, 192.168.2.1\textrm{:}22022, \textrm{SSH}, (5, 5, 5),\\
&(10.0.0.3, 8080,192.168.2.1,3306,TCP),\varnothing )
\end{align*} 

\subsection{Synthetic network scenario generation}\label{sec:net_gen}
All the tests presented in this paper were conducted on a set of synthetically generated networks. 
Note that the generated scenarios do not provide a fully specified network. The output network graph just includes the minimum information used by our anomaly classification algorithms to work (e.g., no middleboxes, not all the connections between network nodes).
Our generation algorithm is parametric and takes as input three arguments:
\begin{enumerate}
	\item the number of non-conflicting \PIs $n_{PI}$;
	\item the number of conflicting \PIs $\overline{n}_{PI}$;
	\item the number of network entities $n_e$.
\end{enumerate}

\begin{figure}[h!]
	\begin{subfigure}[b]{0.5\textwidth}
		\centering
		\begin{tikzpicture}
		\node (a1) {};
		\node[right of = a1] (a2) {};
		\draw[semithick, gray, densely dashed] (a1) -- (a2) node[midway, above, text = Black] (l1) {\small unprotected};
		\node[right of = a2] (b1) {};
		\node[right of = b1] (b2) {};
		\draw[semithick, gray] (b1) -- (b2) node[midway, above, text = Black] (l2) {\small protected};
		\node[right of = b2] (c1) {};
		\node[right of = c1] (c2) {};
		\draw[semithick, gray, double] (c1) -- (c2) node[midway, above, text = Black] (l3) {\small protected tunnel};
		\node [draw = Gray, fit = (a1) (a2) (b1) (b2) (c1) (c2) (l1) (l2) (l3), rounded corners] {};
		\end{tikzpicture}
	\end{subfigure}
	\begin{subfigure}[b]{0.5\textwidth}
		\centering
		\begin{tikzpicture}
		\clientA{client}{};
		\serverA{server}{right of = client, xshift = 10mm};
		\pA{client}{server}{};
		\end{tikzpicture}
		\caption{End-to-end.}
		\label{fig:endToEndScheme}
	\end{subfigure}
	\begin{subfigure}[b]{0.5\textwidth}
		\centering
		\begin{tikzpicture}
		\clientA{client}{};
		\routerA{gw_1}{right of = client, xshift = 5mm};
		\routerA{gw_2}{right of = gw_1, xshift = 5mm};
		\serverA{server}{right of = gw_2, xshift = 5mm};
		\pA{client}{gw_1}{densely dashed};
		\pA{gw_1}{gw_2}{double};
		\pA{gw_2}{server}{densely dashed};
		\end{tikzpicture}
		\caption{Site-to-site with 2 gateways.}
		\label{fig:siteToSiteScheme}
	\end{subfigure}
	\begin{subfigure}[b]{0.5\textwidth}
		\centering
		\begin{tikzpicture}
		\client{client}{};
		\router{gw_1}{right of = client, xshift = 5mm};
		\server{server}{right of = gw_1, xshift = 5mm};
		\pA{client}{gw_1}{double};
		\pA{gw_1}{server}{densely dashed};
		\end{tikzpicture}
		\caption{Remote-access with 1 gateway.}
		\label{fig:remoteAccessScheme}
	\end{subfigure}
	\caption{Network generation schemes.}
\end{figure}
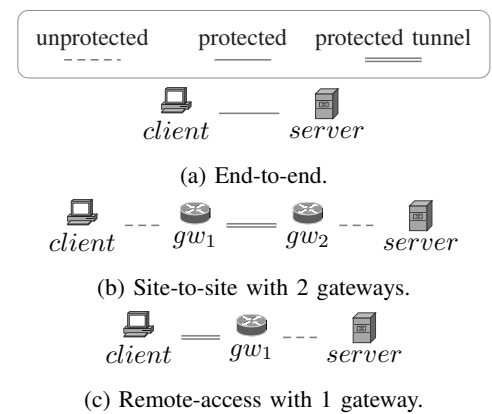

Our approach works in three sequential steps:
\begin{enumerate}
	\item generation of $n_{PI}$ non-conflicting \PIs.
	This phase randomly generates \policyImplementations until the desired number of \PIs is produced. Depending on the type of secure communication types, three different procedures are implemented:
	\begin{itemize}
		\item end-to-end scheme (Fig.~\ref{fig:endToEndScheme}): a new client and a new server are randomly generated and added to the network graph. The connection endpoints are then used to form a single \PI that connects them with randomly chosen non-NULL end to end technology;
		\item site-to-site scheme (Fig.~\ref{fig:siteToSiteScheme}): a new client, a new server and $n_g \ge 2$ gateways are randomly created and added to the network graph. A set of edges are added to the network graph to properly connect the client and the server with the close gateways and to connect the gateways in the proper order. Then, the algorithm creates a \PI connecting the client to the server (which may use the NULL technology) and $n_g - 1$ \PIs to form $n_g - 1$ tunnels that connect pair of adjacent gateways.  The selector of the intended client-server traffic was associated to the \PI;
		\item remote-access scheme (Fig.~\ref{fig:remoteAccessScheme}):  a new client, a new server, and $n_h \ge 1$ gateways are randomly created and added to the network graph nodes. Then, the algorithm creates a \PI connecting the client to the server (which may use the NULL technology) and $n_g$ \PIs  to form $n_g$ tunnels from the client to the last gateway. The selector of the intended client-server traffic was associated to the \PI.
	\end{itemize}

	\item generation of $\overline{n}_{PI}$ conflicting \PIs. This phase randomly generates the \policyImplementations to introduce the anomalies. The algorithm first picks with the same probability one of the nineteen anomalies in our model. Then it selects zero or more network entities generated during the previous step (the exact number depends on the chosen anomaly) and, if needed, it randomly generates the required entities. Finally, it creates the \PIs accordingly;
	\item network graph completion. During this phase, the algorithm first randomly generates the network entities until the limit $n_e$ is reached. Note that this phase might be skipped if all the previous phases have already generated all the required network entities. Subsequently, the algorithm completes the network adding a bare minimum set of network connections, if necessary. Note that, in the general case, full network connectivity is not required, and hence not computed, since we are only interested in computing the analysis time of our implementation and adding such additional data do not impact our tool's performance tests.
\end{enumerate}

\subsection{Notations}\label{sec:notations}

Table~\ref{tab:Notations} lists all the symbols used in this paper.

\begin{table}[h]
	\centering
	\resizebox{\columnwidth}{!}{%
		\begin{tabular}{ll}
			\multicolumn{2}{c}{\textbf{policy implementation}  }\\
			\midrule 
			\multicolumn{1}{l}{	$	i = (s, d, t, C, S, G)$}& \multicolumn{1}{l}{policy implementation (\PI)}\\
			\multicolumn{1}{l}{	$s$}& \multicolumn{1}{l}{channel source}\\
			\multicolumn{1}{l}{$d$}& \multicolumn{1}{l}{channel destination}\\
			\multicolumn{1}{l}{	$t$}& \multicolumn{1}{l}{security technology }\\
			\multicolumn{1}{l}{	$C$}& \multicolumn{1}{l}{required security levels}\\
			\multicolumn{1}{l}{	$S$}& \multicolumn{1}{l}{traffic selectors}\\
			\multicolumn{1}{l}{$G$}& \multicolumn{1}{l}{crossed gateways}\\
			\\
			\multicolumn{2}{c}{\textbf{auxiliary notations}}\\
			\cmidrule(lr){1-2} 
			\multicolumn{1}{l}{$e$}   & \multicolumn{1}{l}{network entity}\\
			\multicolumn{1}{l}{$\overleftarrow{S}$} & \multicolumn{1}{l}{reverse list of selectors} \\
			\multicolumn{1}{l}{$S|_{f_1 \times f_2 \times \dots}$} &  \multicolumn{1}{l}{restrict the selector space} \\
			\multicolumn{1}{l}{$G^*$ }& \multicolumn{1}{l}{crossed gateways with end-points}\\
			\multicolumn{1}{l}{$\overline{G}$ }& \multicolumn{1}{l}{crossed gateways in reverse order}\\
			\multicolumn{1}{l}{	$P^{a, b}$} & \multicolumn{1}{l}{path from $a$ to $b$}\\
			
			\\		 			
			\multicolumn{2}{c}{\textbf{relationships}}\\
			\cmidrule(lr){1-2} 
			$=$& equivalence\\
			$\succ$	&dominance\\
			$\sim$	&kinship\\
			$\perp$	&disjointness\\
			$\not\perp$	&not-disjointness ( \ie $\neq$, $\not\succ$, $\not\sim$)\\
			
			\\		 			
			\multicolumn{2}{c}{\textbf{auxiliary functions}}\\
			\cmidrule(lr){1-2} 
			\multicolumn{1}{l}{$\mathcal{N}(i)$ } &\multicolumn{1}{l}{node where the \PI $i$ is deployed}\\
			\multicolumn{1}{l}{$\mathcal{T}(e)$ }	& \multicolumn{1}{l}{technologies supported by node $e$}\\
			\multicolumn{1}{l}{$\mathcal{C}_{max}(i)$ }& \multicolumn{1}{l}{maximum coefficients supported by the \PI $i$}\\
			\multicolumn{1}{l}{	$\mathcal{C}_{min}(i)$ }& \multicolumn{1}{l}{minimum coefficients acceptable for the \PI $i$}\\
			\multicolumn{1}{l}{	$\pi(i)$ }& \multicolumn{1}{l}{priority of the \PI $i$}\\
			\multicolumn{1}{l}{$\mathcal{F}_e(i)$ }& \multicolumn{1}{l}{filtering of node $e$ for the \PI $i$}\\
			\multicolumn{1}{l}{$\mathcal{T}^{(2)}(e)$} & \multicolumn{1}{l}{technologies supported at level 2 for the node $e$}\\
		\end{tabular}
	}
	\caption{Mathematical notations.}
	\label{tab:Notations}
\end{table}

\end{document}